\documentclass[preprint,3p,times]{elsarticle}

\biboptions{sort&compress}

\usepackage{amsmath}
\usepackage{amssymb}
\usepackage{booktabs}
\usepackage{graphicx}
\usepackage{hyperref}
\usepackage{doi}
\usepackage{pdflscape}
\usepackage{afterpage}
\usepackage{listings}
\usepackage{cleveref}
\journal{Computer Physics Communications}

\begin{document}

\begin{frontmatter}

\title{libwignernj: a reusable C/C++/Fortran/Python library for
  exact Wigner symbols and related coefficients}

\author[orcid=0000-0001-6296-8103]{Susi Lehtola}
\ead{susi.lehtola@helsinki.fi}
\affiliation{organization={Department of Chemistry, University of Helsinki},
             addressline={P.\,O.\,Box 55},
             postcode={FIN-00014},
             country={Finland}}

\begin{graphicalabstract}
\centering
\includegraphics[width=0.98\textwidth]{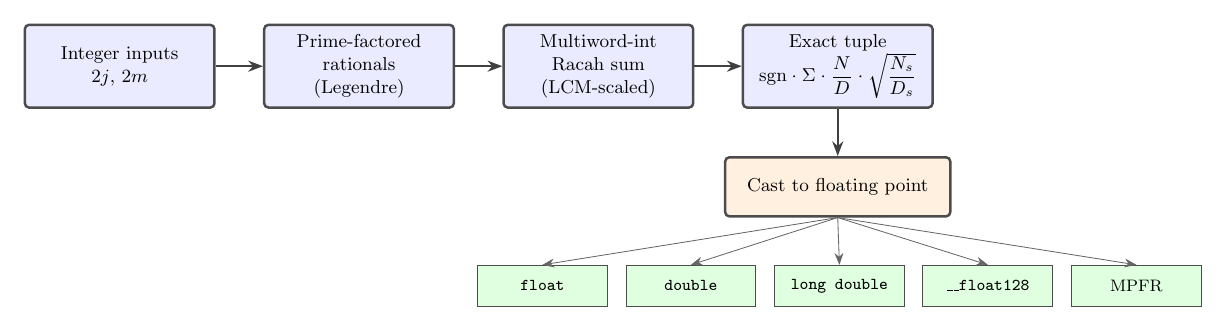}
\end{graphicalabstract}

\begin{highlights}
\item Exact prime-factorisation pipeline for Wigner 3j/6j/9j, CG, W, and Gaunt symbols
\item Last-bit-correct results at float, double, long double, libquadmath, and MPFR precision
\item C, C++, Fortran 90, and Python bindings. No caller-side initialisation
\item Dedicated routines for complex and real-spherical-harmonic Gaunt coefficients
\item BSD-licensed; CMake- and pkg-config-installable; CI on Linux, macOS and Windows
\end{highlights}

\begin{abstract}
We describe \texttt{libwignernj}, a freely available, BSD-licensed
library that evaluates Wigner 3j, 6j, and 9j symbols,
Clebsch--Gordan, Racah $W$, and Fano $X$ coefficients, and Gaunt
coefficients over both complex and real spherical harmonics in
standards-compliant C99. \texttt{libwignernj} represents factorials
by the vector of their signed prime-exponent decomposition---a
prime-factorization technique introduced for the angular-momentum
coefficients by Dodds and Wiechers (Comput.\ Phys.\ Commun.\ 4, 268
(1972)) and refined in a long line of subsequent work---and combines
that representation with the multiword-integer Racah sum of Johansson
and Forss\'en (SIAM J.\ Sci.\ Comput.\ 38, A376 (2016)), under which
every intermediate quantity is an exact rational and all rounding is
confined to the final floating-point conversion. Single-, double-,
and long-double-precision results are correct to the last
representable bit, and IEEE~754 binary128 evaluation through
libquadmath and arbitrary-precision evaluation through the GNU
Multiple-Precision Floating-Point Reliable (MPFR) library are
optionally exposed. \texttt{libwignernj} has no mandatory runtime
dependencies and no caller-side initialization step, making it easy
to embed across the atomic, molecular, nuclear, and
electromagnetic-scattering applications in which these coefficients
arise. C\texttt{++}, CPython, and Fortran~90 bindings ship alongside
the C library. Half-integer angular momenta are encoded exactly via
integer $2j$ arguments throughout the application programming
interface (API).  CMake-package and pkg-config files ship for
drop-in integration into downstream projects, and a continuous-%
integration (CI) pipeline runs the full test suite on Linux (shared
and static), macOS, and Windows on every push.
\end{abstract}


\end{frontmatter}

%--------------------------------------------------------------------
\section*{Program summary}
\noindent\begin{tabular}{@{}p{0.32\textwidth}p{0.62\textwidth}@{}}
\hline
\textit{Program title}            & \texttt{libwignernj} \\
\textit{Licensing provisions}     & BSD 3-Clause \\
\textit{Code repository}          & \url{https://github.com/susilehtola/libwignernj} \\
\textit{CPC Library link}         & \url{https://doi.org/10.17632/5t9r7zytbm.2} \\
\textit{Programming language}     & C99 (core); C++11, Fortran~90, and Python~3 (interfaces) \\
\textit{External dependencies}    & None for the core library; libquadmath
                                    (shipped with GCC) is required for the optional binary128 back-end; GNU MPFR is required for the optional arbitrary-precision back-end \cite{Fousse2007_ATMS_13}; FLINT~\cite{Hart2010__88} (in turn pulling in GMP and GNU MPFR) is required for the optional sub-quadratic bigint back-end; \texttt{setuptools} for the Python build \\
\textit{Operating systems}        & Linux, macOS, Microsoft
                                    Windows, and any other target with a C99 compiler \\
\textit{Build system}             & CMake $\ge\,3.16$, with optional
                                    out-of-tree Python build via \texttt{pip install -e .} \\
\textit{Nature of physical problem} & Exact, last-bit-correct
                                    floating-point evaluation of Wigner 3j, 6j and 9j symbols, Clebsch--Gordan coefficients, Racah $W$ coefficients, Fano $X$-coefficients, and Gaunt coefficients (the angular integral of three spherical harmonics), at single, double, long-double, IEEE~754 binary128 (libquadmath), or arbitrary precision \\
\textit{Method of solution}       & Prime-factorization of the Racah
                                    sum~\cite{Johansson2016_SJSC_376}: each factorial is represented by its vector of prime exponents (Legendre's formula), and each Racah-sum term is converted to a multiword integer through a per-symbol least-common-multiple denominator. The single floating-point step is the bigint-to-float cast at the end \\
\textit{Restrictions}             & Default-build ceiling at
                                    $j_1+j_2+j_3 \le 20019$ for 3j, 6j, CG, Racah $W$, complex Gaunt, and real-spherical-harmonic Gaunt, and equal-$j$ ceiling $j \le 5004$ for 9j and Fano $X$ (\cref{sec:limits}); set by the size of the compile-time prime table and raised by regenerating it with a larger sieve limit. The per-symbol cost scales as $O(j^2)$ for 3j and 6j and $O(j^4)$ for 9j and Fano $X$ \\
\textit{Typical running time}     & 3j: $\sim 1\,\mu$s at $j\sim 5$,
                                    $\sim 0.5\,$ms at $j\sim 1000$; 6j: $\sim 3\,\mu$s at $j\sim 5$, $\sim 10\,$ms at $j\sim 500$; 9j: tens of microseconds at $j\sim 5$, $\sim 100\,$ms at $j\sim 80$, on a single x86-64 core (see \cref{sec:benchmarks} for representative timings and comparison with WIGXJPF and the GNU Scientific Library, GSL) \\
\hline
\end{tabular}

\section{Introduction}

The angular-momentum coupling coefficients of quantum
mechanics---the Clebsch--Gordan (CG) coefficients and Wigner
3j/6j/9j symbols
of~\cite{Wigner1993__608,Racah1942_PR_186,Racah1942_PR_438,Racah1943_PR_367,Racah1949_PR_1352,Jahn1954_PR_318},
the closely-related Racah $W$
coefficient~\cite{Racah1942_PR_438,Racah1943_PR_367}, and the
Gaunt integral over three spherical
harmonics~\cite{Gaunt1929_PTRSAMPES_151}---enter virtually every
numerical calculation that involves rotational symmetry.  They
appear in the Wigner--Eckart matrix elements that determine line
strengths, multiplet splittings, and hyperfine couplings in
atomic structure~\cite{Johnson2007__,Friedrich2017__}. They dominate
two-body matrix elements in nuclear shell-model and \emph{ab
initio} no-core-shell-model calculations, where modern
interactions can require many millions of 6j and 9j evaluations
per matrix construction~\cite{Caurier2005_RMP_427}. In molecular
electronic-structure theory the Gaunt coefficient appears in the
multipole resolution of the inter-electron Coulomb interaction
over real spherical-harmonic basis sets~\cite{Helgaker2000__}.
In classical electromagnetic scattering it enters the
addition theorem for vector spherical harmonics underlying
multi-particle Mie and $T$-matrix algorithms~\cite{Xu1996_MC_1601}.
The need for fast and reliable evaluation across these
applications has driven a long line of published
implementations, which we collect in \cref{tab:prior-work} and
survey below by method.

\paragraph{Direct floating-point evaluation}
The most straightforward strategy evaluates the Racah single-sum
formula directly in floating-point arithmetic.  \citet{Caswell1966__} distributed the earliest Fortran
programs (3j, 6j, 9j up to $j\le 80$) as a National Bureau of
Standards Technical Note, with the 9j as a sum of 6j products.
\citet{Tamura1970_CPC_337} contributed the first
Computer Physics Communications (CPC) version, pre-tabulating
$\log n!$ at start-up.  \citet{Wills1971_CPC_381} reformulated
Tamura's CG inner-loop sum as a Horner-style nested product to
remove the factorial-overflow risk, and
\citet{Bretz1976_APH_255} extended the Horner reformulation
to the 6j.  Two later families recast the inner sum to avoid
factorials entirely: \citet{SrinivasaRao1978_CPC_227,SrinivasaRao1981_CPC_297}
as generalised hypergeometric functions ${}_3F_2$ and ${}_4F_3$
evaluated by Horner's rule, and \citet{Guseinov1995_JCP_343,Guseinov2009_JTCC_251} as
binomial-coefficient sums in REAL*16, with
\citet{Wei1998_CP_632} giving the binomial single-sum 9j.
\citet{Shriner1993_CP_144} reviewed the field
in a contemporaneous article.

\paragraph{Recursive evaluation}
\citet{Schulten1975_JMP_1961,Schulten1976_CPC_269,Schulten1984_CPC_377}
replaced the Racah sum with three-term recursions iterated
outwards from both classical turning points and matched in the
classical region.  \citet{Luscombe1998_PRE_7274} subsequently simplified its
bookkeeping.  The same idea has been applied to other families:
\citet{Xu1996_MC_1601,Xu1997_JCAM_53} reduced the Gaunt
coefficient to a single recursion accurate at both low and high
degree, and more recently \citet{Xu2020_JQSRT_107210} gave a
Clebsch--Gordan recursion targeting radiative-transfer
applications.

\paragraph{Symbolic / graphical reduction}
\citet{Burke1970_CPC_241,Burke1984_CPC_30} introduced
\texttt{NJSYM} for general 3$n$-$j$ recoupling coefficients via
the graphical reduction of recoupling
networks~\cite{Yutsis1962__,Massot1967_RMP_288}\footnote{In Burke's
terminology, ``$n$-$j$ symbols'' meant the same family that later
authors call ``3$n$-$j$ symbols''---the prefactor of three was
implicit.  The first three cases $n=1,2,3$ are the 3j, 6j, and 9j
symbols, with $3n$ angular-momentum labels in all (the $n+1$
external momenta plus the intermediate-sum labels of each scheme).
We adopt the modern 3$n$-$j$ convention throughout this paper.}.
\citet{Shapiro1970_CPC_207,Shapiro1984_CPC_25} gave a
related Fortran code for arbitrary 3$n$-$j$ symbols of SU(2).
\citet{Scott1982_CPC_189} provided more efficient
versions of \texttt{WEIGHTS} and \texttt{NJSYM}.  \citet{BarShalom1988_CPC_375} subsequently replaced
\texttt{NJSYM} with \texttt{NJGRAF}, which generates an optimised
sum-over-6j-products and then evaluates the 6j's numerically (up
to two orders of magnitude faster at unchanged precision).  \citet{Fack1997_CPC_155} continued with
\texttt{GYutsis}, a graph-rewriting program for arbitrary 3$n$-$j$
recoupling, and later showed that minimising rotation distance on
the binary coupling tree reduces the resulting sum
substantially~\cite{Fack1999_CPC_99}.

On the fully-symbolic side, \citet{Koike1992_CPC_154}
provided \texttt{ANALG}, a \texttt{REDUCE} program for algebraic
reductions of CG, 6j, 9j, and $X$-coefficient formulas.
\citet{Stevenson2002_CPC_853} contributed a Java applet that
returns coefficients in their analytical
``rational~$\times \sqrt{\text{rational}}$'' form.
\citet{Fritzsche1997_CPC_51,Fritzsche2009_CPC_2021} developed the
\texttt{RACAH} Maple package, with extensions covering recoupling,
sum-rule reduction, and many-particle matrix elements.
\citet{Deveikis2005_CPC_60} provided a Scheme calculator that
exploits Scheme bignums for exact evaluation, and
\citet{Xiang2021_CPC_107880} a Python program that simplifies
sums of Wigner 3$j$-symbols via
Yutsis--Levinson--Vanagas~\cite{Yutsis1962__,Massot1967_RMP_288}
graphical techniques.  Beyond these dedicated programs,
mainstream computer algebra systems implement the same
coefficients out of the box: Mathematica's built-in
\texttt{ThreeJSymbol}, \texttt{SixJSymbol}, and
\texttt{ClebschGordan} (with \texttt{NineJSymbol} available in the
Wolfram Function Repository); Maxima's \texttt{clebsch\_gordan}
contrib package, which exposes \texttt{wigner\_3j},
\texttt{wigner\_6j}, and \texttt{wigner\_9j}; and
\texttt{sympy.\allowbreak physics.\allowbreak wigner}~\cite{Meurer2017_PCS_103}.

\paragraph{Parallel evaluation}
Two CPC programs explored parallelisation rather than the
underlying algorithm: \citet{Scott1987_CPC_83} distributed the inner Racah-sum loop
of the 6j across an array of transputers in Occam, and Fack
\emph{et~al.}~\cite{Fack1992_CPC_285} extended the same approach
to the 9j (via the sum-over-three-6j formula) in Parallel~C.

\paragraph{Exact integer / rational arithmetic}
The earliest evaluation of 3-$j$ and 9-$j$ symbols using
multiprecision rational arithmetic appears to be that of
\citet{Baer1964_CA_657}. \citet{SrinivasaRao1989_CPC_231} gave a
9$j$-specific Fortran program in which each intermediate factor of
the sum-over-three-6j-products is stored as an arbitrary-precision
integer. \citet{Tuzun1998_CPC_112} rewrote the 3j and 6j formulas
as alternating binomial sums and observed that each binomial sum
is itself an integer---representable as a double for moderate
angular momenta and as a prime-factored integer otherwise---while
the prefactors are evaluated separately without round-off. The
same paper analysed the loss-of-significance behaviour of direct
double-precision evaluation in detail.

\paragraph{Prime factorization}
Representing each integer factor in terms of its prime
decomposition allows turning multiplications and divisions of
large factorials into additions and subtractions of small signed
integers, with a dynamic range that never exceeds that of a
$\log_2$-sized exponent.  This approach has a long lineage in the
coupling-coefficient literature.  \citet{Dodds1972_CPC_268} introduced it for the 3j and 6j
symbols.  \citet{Stone1980_CPC_195} generalised it to
a list-processing Fortran package covering the 9j symbol; Fang
and
\citet{Fang1992_CPC_147} extended the arithmetic to CG,
Racah, and Wigner symbols, returning $(a/b)\sqrt{c/d}$ explicitly.  \citet{Lai1990_CPC_350,Lai1992_CPC_544} kept the
prefactor exact-integer but evaluated the alternating sum in
REAL*16.  \citet{Wei1999_CPC_222,Wei2011_CPC_1199} combined
prime-factor prefactors with a base-32768 multiword back-end for
the binomial-coefficient summation.  The decisive step was made by
\citet{Johansson2016_SJSC_376}, who combined prime factorisation
with multiword-integer accumulation of the alternating Racah sum.
This way each prime is touched at most $\log_p N$ times
(Legendre's formula), and the only floating-point rounding occurs
in the final cast.  Their \texttt{WIGXJPF} achieved bit-exact
evaluation faster than any prior implementation.

\paragraph{Storage and lookup schemes}
Orthogonal to the choice of evaluation algorithm, several authors
have addressed the distinct problem of storing pre-computed
coefficients efficiently for repeated lookup, a useful primitive
when the same coefficients are needed many times in an outer loop.
\citet{Vermaak1984_CPC_41} described an early packed storage
scheme for the 3j symbol exploiting its symmetries.  \citet{Rasch2004_SJSC_1416} extended the philosophy to the 6j and
Gaunt coefficients with carefully designed index functions
demonstrating an order-of-magnitude speedup over recomputation.  \citet{Pinchon2007_IJQC_2186} introduced even more compact index
functions for the (almost-all-nonzero) Gaunt coefficients of
molecular electronic-structure calculations.
\citet{Guseinov2005_JMST_177} gave a shared-storage scheme for
Clebsch--Gordan and Gaunt coefficients exploiting their common
selection rules.

\paragraph{Gaunt-coefficient-specific work}
Beyond the direct and recursive Gaunt evaluations cited
above~\cite{Guseinov1995_JCP_343,Xu1996_MC_1601,Guseinov2009_JTCC_251},
\citet{Homeier1996_JMST_31} treated the
closely related coupling coefficients of \emph{real} spherical
harmonics. Recent work has continued along four directions: new
explicit representations and orthogonality
relations~\cite{Yuekcue2019_CPL_136769,Oezay2024_CPC_109118},
recurrences for the real-spherical-harmonic
variant~\cite{Yuekcue2025_CJP_321}, a re-derivation in the
conventions of spherical array processing and
\citet{Politis2024__} with an accompanying
\texttt{MATLAB} implementation, and unified evaluation of the
Gaunt, CG, 3$j$, and 6$j$ coefficients as products of
generalised hypergeometric functions by
\citet{Oezay2025_CPC_109656}.

\paragraph{Modern open-source libraries}
In addition to the peer-reviewed implementations surveyed above,
a complementary set of open-source coupling-coefficient libraries
is distributed on code-hosting platforms.  They are collected in
\cref{tab:oss}.  The Johansson--Forss\'en prime-factorisation
pipeline has been ported to Julia
(\texttt{WignerSymbols.jl}~\cite{Haegeman_WignerSymbolsjl}) and to
Rust (\texttt{wigners}~\cite{Fraux_wigners}). The Wei
binomial-coefficient route is taken by
\texttt{CGcoefficient.jl}~\cite{0382_CGcoefficientjl}, which
additionally covers Moshinsky brackets~\cite{Moshinsky1959_NP_104} (M.b.). The Schulten--Gordon and
Luscombe--Luban recursions are exposed by
\texttt{wignerSymbols}~\cite{Dumont_wignerSymbols},
\texttt{py3nj}~\cite{Fujii_py3nj} (vectorised Python wrapping
\texttt{SLATEC}), and
\texttt{WignerFamilies.jl}~\cite{Li_WignerFamiliesjl}, while
\texttt{spherical}~\cite{Boyle_spherical} provides a
recursion-with-caching variant in Python/Numba widely used in the
gravitational-wave community. \texttt{wigner}~\cite{Gorton_wigner}
is a modern-Fortran direct-evaluation implementation, and
\texttt{FASTWIGXJ}~\cite{Johansson_FASTWIGXJ} is the
\texttt{WIGXJPF} authors' pre-tabulation and dynamic-hashing
companion.

\afterpage{%
\clearpage
\begin{landscape}
\begin{table}[t]
\centering
\caption{Peer-reviewed implementations of angular-momentum coupling
  coefficients referenced in the present work. ``CG'' denotes
  Clebsch--Gordan; ``$W$'' denotes Racah $W$. Symbol-set entries are
  what is computed by the cited code, not what the cited paper
  derives in theory. CPC library re-issues are cited next to the
  original paper where relevant.}
\label{tab:prior-work}
\footnotesize
\begin{tabular}{@{}lllll@{}}
\toprule
Reference & Coefficients & Language & Comment & DOI \\
\midrule
\citet{Baer1964_CA_657}              & 3j, 9j           & ALGOL 60   & multiprecision rational arithmetic                & \doi{10.1145/364984.365075} \\
\citet{Caswell1966__}              & 3j, 6j, 9j       & Fortran    & direct evaluation; NBS Technical Note 409          & \doi{10.6028/NBS.TN.409} \\
\citet{Burke1970_CPC_241}                              & 3$n$-$j$         & Fortran    & graphical reduction to sum of 6j products; \texttt{NJSYM} & \doi{10.1016/0010-4655(70)90040-8} \\
\citet{Shapiro1970_CPC_207,Shapiro1984_CPC_25}              & 3$n$-$j$         & Fortran    & graphical reduction (SU(2))                       & \doi{10.1016/0010-4655(70)90007-X} \\
\citet{Tamura1970_CPC_337}                            & CG, 6j, 9j       & Fortran    & direct evaluation via stored log-factorial table  & \doi{10.1016/0010-4655(70)90034-2} \\
\citet{Wills1971_CPC_381}                              & CG               & Fortran    & Horner-product reformulation of Tamura's CG sum   & \doi{10.1016/0010-4655(71)90030-0} \\
\citet{Dodds1972_CPC_268}          & 3j, 6j           & Fortran    & prime-factor representation of integer arithmetic & \doi{10.1016/0010-4655(72)90019-7} \\
\citet{Bretz1976_APH_255}                              & 6j, CG           & ?          & extension of Wills's Horner-product method to 6j  & \doi{10.1007/BF03157502} \\
\citet{Schulten1976_CPC_269,Schulten1984_CPC_377} & 3j, 6j & Fortran    & three-term recursion                            & \doi{10.1016/0010-4655(76)90058-8} \\
\citet{SrinivasaRao1978_CPC_227} & CG, $W$ ($=$~6j) & Fortran    & hypergeometric ${}_3F_2$/${}_4F_3$ forms          & \doi{10.1016/0010-4655(78)90093-0} \\
\citet{Stone1980_CPC_195}                  & 3j, 6j, 9j       & Fortran    & power-of-prime arithmetic via list-processing     & \doi{10.1016/0010-4655(80)90040-5} \\
\citet{SrinivasaRao1981_CPC_297}               & 3j, 6j           & Fortran    & generalized hypergeometric functions              & \doi{10.1016/0010-4655(81)90063-1} \\
\citet{Scott1982_CPC_189}            & 3$n$-$j$         & Fortran    & efficient \texttt{WEIGHTS}/\texttt{NJSYM}          & \doi{10.1016/0010-4655(82)90054-6} \\
\citet{Vermaak1984_CPC_41}            & 3j               & Fortran    & packed storage of pre-tabulated values             & \doi{10.1016/0010-4655(84)90080-8} \\
\citet{Scott1987_CPC_83}              & Racah ($=$~6j)   & Occam      & parallel transputer evaluation                     & \doi{10.1016/0010-4655(87)90037-3} \\
\citet{BarShalom1988_CPC_375} & 3$n$-$j$         & Fortran    & graphical analysis $\to$ minimal sum-over-6j-products; \texttt{NJGRAF} & \doi{10.1016/0010-4655(88)90192-0} \\
\citet{SrinivasaRao1989_CPC_231} & 9j           & Fortran    & exact integer arithmetic                            & \doi{10.1016/0010-4655(89)90021-0} \\
\citet{Lai1990_CPC_350}                      & 3j, 6j           & Fortran    & prime decomposition of prefactor + REAL*16 sum    & \doi{10.1016/0010-4655(90)90049-7} \\
\citet{Fack1992_CPC_285}              & 9j (recoupling)  & Parallel~C & parallel transputer evaluation                     & \doi{10.1016/0010-4655(92)90015-Q} \\
\citet{Fang1992_CPC_147}              & 3j, 6j, 9j, CG, $W$ & Fortran & prime-factor arrays giving $(a/b)\sqrt{c/d}$       & \doi{10.1016/0010-4655(92)90097-I} \\
\citet{Koike1992_CPC_154}                              & CG, 3j, 6j, 9j, $X$ & REDUCE  & symbolic algebraic reduction                       & \doi{10.1016/0010-4655(92)90147-Q} \\
\citet{Lai1992_CPC_544}                      & 9j               & Fortran    & prime decomposition of prefactor + REAL*16 sum    & \doi{10.1016/0010-4655(92)90115-F} \\
\citet{Guseinov1995_JCP_343}          & CG, Gaunt        & Fortran    & binomial-coefficient direct sum (REAL*16)         & \doi{10.1006/jcph.1995.1220} \\
\citet{Xu1996_MC_1601}                                    & Gaunt            & ?          & lower-triangular linear system + recurrences      & \doi{10.1090/S0025-5718-96-00774-0} \\
\citet{Fack1997_CPC_155}              & 3$n$-$j$         & C          & graph-rewriting recoupling reduction; \texttt{GYutsis} & \doi{10.1016/S0010-4655(96)00170-1} \\
\citet{Fritzsche1997_CPC_51,Fritzsche2009_CPC_2021} & CG, 3j, 6j, 9j, recoupling & Maple & symbolic Racah algebra; \texttt{RACAH} & \doi{10.1016/S0010-4655(97)00032-5} \\
\citet{Xu1997_JCAM_53}                                    & Gaunt            & ?          & single-index recursion (low and high degree)       & \doi{10.1016/S0377-0427(97)00128-3} \\
\citet{Tuzun1998_CPC_112}& 3j, 6j           & Fortran    & alternating binomial sums (FP or prime-factored int) & \doi{10.1016/S0010-4655(98)00065-4} \\
\citet{Wei1998_CP_632}                                 & 9j               & ?          & single-sum binomial-coefficient direct evaluation (double precision)  & \doi{10.1063/1.168745} \\
\citet{Wei1999_CPC_222,Wei2011_CPC_1199}              & 3j, 6j, 9j       & Fortran    & prime-factor prefactor + base-32768 multiword sum & \doi{10.1016/S0010-4655(99)00232-5} \\
\citet{Stevenson2002_CPC_853}              & CG, 3j, 6j, 9j   & Java       & arbitrary-precision exact analytical evaluation     & \doi{10.1016/S0010-4655(02)00462-9} \\
\citet{Rasch2004_SJSC_1416}                      & 3j, 6j, Gaunt    & C, Fortran & efficient storage of pre-tabulated values          & \doi{10.1137/S1064827503422932} \\
\citet{Deveikis2005_CPC_60}& CG, 6j, 9j, recoupling   & Scheme     & direct evaluation via Scheme bignums              & \doi{10.1016/j.cpc.2005.06.003} \\
\citet{Guseinov2005_JMST_177}      & CG, Gaunt        & Turbo Pascal & shared-storage scheme via common selection rules & \doi{10.1016/j.theochem.2004.08.036} \\
\citet{Pinchon2007_IJQC_2186}          & Gaunt            & C          & compact index functions for storage                & \doi{10.1002/qua.21337} \\
\citet{Guseinov2009_JTCC_251}          & CG, Gaunt, $n$-$j$ & ?        & extended binomial-coefficient method             & \doi{10.1142/S0219633609004782} \\
\citet{Johansson2016_SJSC_376}  & 3j, 6j, 9j       & C          & prime factorization + multiword integer Racah sum  & \doi{10.1137/15M1021908} \\
Y{\"u}k\c{c}{\"u} \emph{et~al.}~\cite{Yuekcue2019_CPL_136769}    & Gaunt            & Mathematica & new explicit representations                      & \doi{10.1016/j.cplett.2019.136769} \\
\citet{Xu2020_JQSRT_107210}                                    & CG               & MATLAB     & improved recursive evaluation                      & \doi{10.1016/j.jqsrt.2020.107210} \\
\citet{Xiang2021_CPC_107880}                  & 3j sums          & Python     & graphical (Yutsis) symbolic simplification          & \doi{10.1016/j.cpc.2021.107880} \\
{\"O}zay \emph{et~al.}~\cite{Oezay2024_CPC_109118}              & Gaunt            & Mathematica & new orthogonality-relation-based evaluation       & \doi{10.1016/j.cpc.2024.109118} \\
Y{\"u}k\c{c}{\"u} \emph{et~al.}~\cite{Yuekcue2025_CJP_321}    & Gaunt (real $Y$) & Mathematica & direct sum / recurrence for real $Y_{\ell m}$    & \doi{10.1139/cjp-2024-0161} \\
{\"O}zay \emph{et~al.}~\cite{Oezay2025_CPC_109656}              & Gaunt, CG, 3j, 6j & Mathematica & generalised hypergeometric form; \texttt{Gaunt\_CG\_3j\_and\_6j} & \doi{10.1016/j.cpc.2025.109656} \\
\bottomrule
\end{tabular}
\end{table}
\end{landscape}
\clearpage
}

\begin{table}[t]
\centering
\caption{Active open-source software libraries for angular-momentum
  coupling coefficients distributed outside the journal literature.}
\label{tab:oss}
\footnotesize
\begin{tabular}{@{}lllll@{}}
\toprule
Library & Coefficients & Language & Method & License \\
\midrule
\texttt{WIGXJPF}                                       & 3j, 6j, 9j               & C            & prime factorisation + multiword int.            & LGPL-3.0 \\
\texttt{FASTWIGXJ}~\cite{Johansson_FASTWIGXJ}          & 3j, 6j, 9j               & C            & tabulation + dynamic hash                & LGPL-3.0 \\
\texttt{WignerSymbols.jl}~\cite{Haegeman_WignerSymbolsjl} & 3j, 6j, CG, $V$, $W$  & Julia        & prime factorisation         & MIT \\
\texttt{wigners}~\cite{Fraux_wigners}                  & 3j, CG                   & Rust, Python & prime factorisation                                 & Apache-2.0 \\
\texttt{CGcoefficient.jl}~\cite{0382_CGcoefficientjl}  & 3j, 6j, 9j, CG, $W$, M.b.& Julia        & Wei binomial-coefficient sum    & MIT \\
\texttt{wignerSymbols}~\cite{Dumont_wignerSymbols}     & 3j, 6j                   & C\texttt{++}/Fortran & Schulten--Gordon recursion                            & LGPL-3.0 \\
\texttt{py3nj}~\cite{Fujii_py3nj}                      & 3j, 6j, 9j, CG           & Py wrap SLATEC & Schulten--Gordon, vectorised               & Apache-2.0 \\
\texttt{WignerFamilies.jl}~\cite{Li_WignerFamiliesjl}  & 3j, 6j (families)        & Julia        & Luscombe--Luban family recurrences     & MIT \\
\texttt{wigner}~\cite{Gorton_wigner}                   & 3j, 6j, 9j               & Fortran      & direct evaluation, OpenMP, lookup tables  & MIT \\
\texttt{spherical}~\cite{Boyle_spherical}              & 3j, $D$-matrix, sYlm     & Python/Numba & recursion + caching                                 & MIT \\
\bottomrule
\end{tabular}
\end{table}

As \Cref{tab:prior-work} and the survey above make clear,
a wide variety of algorithmic approaches has been pursued in the
literature, ranging from direct floating-point evaluation of the
Racah single sum, through three-term recursions and graphical
reduction, to symbolic algebra and exact-integer arithmetic.
This survey enumerates the published implementations by their
stated method and scope; it does not constitute an independent
audit of their internal consistency, numerical reliability, or
correctness, nor does the inclusion of a citation in
\cref{tab:prior-work} amount to an endorsement of the
underlying approach.  We have benchmarked \texttt{libwignernj}
against only the two reference implementations relevant to its
own design choices, \texttt{WIGXJPF} and the GNU Scientific
Library (\cref{sec:benchmarks}); a broader comparative study
across the body of published implementations---several of
which target different coefficient families, different host
languages, or different accuracy regimes than those addressed
here---would be a substantial undertaking in its own right and
falls outside the scope of the present work. The primary
interest here is the subset of approaches that eliminate
finite-precision errors from the intermediate computation, so
that the same algorithm delivers correctly-rounded results in
any chosen floating-point precision with no algorithmic
redesign.  Among the published approaches, the prime-factorisation pipeline of
\citet{Johansson2016_SJSC_376} stands out, as it touches each
prime factor only $\log_p N$ times and confines all rounding to
a single final cast.  The present work extends the same
exact-arithmetic treatment to coefficients the original paper
did not consider---in particular the Clebsch--Gordan and Racah
$W$ coefficients and the Gaunt coefficients over both complex
and real spherical harmonics.

Beyond the algorithmic question of fast and accurate
evaluation, the present work is also shaped by an epistemic
aim. As I have already argued
elsewhere~\cite{Lehtola2023_JCP_180901}, modern computational
science depends critically on its software infrastructure, and
that infrastructure is most effective when delivered as small,
focused, reusable open-source libraries with clear interfaces
and permissive licences, rather than as features embedded
inside a single large code base. Reproducibility imposes the
same requirement: an in-principle reproducible calculation is
reproducible in practice only if the entire software stack
underneath it is open-source and inspectable, so that an
independent reader can audit, recompile, and re-run it on their
own hardware---closed-source or licence-restricted dependencies
leave gaps in this chain of evidence that no external user can
close~\cite{Lehtola2022_WIRCMS_1610}. We have also recently
discussed these issues in the specific cases of the
reproducible implementation of density
functionals~\cite{Lehtola2023_JCP_114116} and of
self-consistent-field solvers and orbital
optimization~\cite{Lehtola2025_JPCA_5651,Greiner2026_JCTC_881}.
Angular-momentum coupling coefficients are a textbook example
of a problem that benefits from this philosophy: every
reasonably sophisticated calculation in atomic, molecular, or
nuclear physics needs them, and every group has historically
rolled their own implementation.  Yet, the API is so simple in
this case that a single shared, stable public API is both
feasible and durable: a small set of input integers maps to a
single correctly-rounded floating-point output, with no
algorithmic choices, convergence parameters, or
domain-specific data structures to expose as in our other
recent reusable libraries.

The concrete aims of the present work can be summarised as:
\begin{enumerate}
\item \textbf{Numerical reliability.} Last-bit-correct results at
  single, double, long-double, and (optional) IEEE~754 binary128
  precision via the prime-factorization scheme, with no silent
  overflow or underflow in intermediate arithmetic.  The public
  API takes integer $2j$ arguments throughout, so half-integer
  angular momenta are encoded exactly without any floating-point
  approximation.
\item \textbf{Permissive licensing.} BSD~3-Clause, so the library
  can be embedded in proprietary or differently-licensed
  scientific software, unlike the copyleft GSL (GPL-3.0) and
  \texttt{WIGXJPF} (LGPL-3.0).
\item \textbf{Reusable-software design with broad language
  coverage} in the spirit of our recent ``call to
  arms''~\cite{Lehtola2023_JCP_180901}, rather than features
  buried inside a monolithic code.  Following the
  architectural template of
  \texttt{Libxc}~\cite{Marques2012_CPC_2272,Lehtola2018_S_1},
  we design a small, focused core library with thin language
  wrappers.  A single source exposes C, C\texttt{++}11,
  Fortran~90 (\texttt{iso\_c\_binding}), and Python (CPython,
  with a \texttt{precision=} keyword); earlier public
  implementations expose at most two of these.
\item \textbf{No external runtime dependencies.} Self-contained
  C99 with its own multiword integer arithmetic; libquadmath,
  GNU~MPFR, and FLINT (Fast Library for Number
  Theory~\cite{Hart2010__88}) are needed only for their
  respective optional back-ends.
\item \textbf{Bundled coefficients.} Clebsch--Gordan, Racah $W$,
  Fano $X$, complex Gaunt, and real-spherical-harmonic Gaunt are
  all first-class entries in the public API, each with its own
  exact-arithmetic pipeline, rather than left to the caller to
  assemble from 3j/6j/9j calls.
\item \textbf{No caller-side initialisation.} The prime table is a
  compile-time constant, so a single function call returns a
  result.  \texttt{WIGXJPF}, in contrast, requires
  \texttt{wig\_table\_init} and \texttt{wig\_temp\_init}
  allocations whose one-time cost grows with the maximum $2j$ that
  will ever be passed (\cref{sec:benchmarks}).  An optional
  \texttt{wignernj\_warmup\_to} entry point is provided for callers
  who want to pre-populate the per-thread caches up front.
\item \textbf{Portability.} The default build uses
  \texttt{\_\_uint128\_t} where available; a pure-C99 fallback
  path is exercised in continuous integration so the library also
  compiles cleanly on Microsoft Visual C\texttt{++} (MSVC) and any
  other compiler that lacks the extension (\cref{sec:limits}).
\item \textbf{Drop-in CMake integration.} The build ships a
  versioned CMake package and a pkg-config file
  (\cref{sec:software}), equally usable as an installed
  dependency or as a git submodule.  To the best of our
  knowledge, none of the earlier public implementations of
  these coefficients ships either of these out of the box.
\item \textbf{Thorough continuous integration.} Every push and
  pull request runs the entire test suite through GitHub Actions
  and CircleCI pipelines that span: Linux/GCC (shared and static
  builds), Linux/Clang, macOS/Clang, and Windows/MSVC; native
  arm64-Linux, musl-libc x86\_64, and 32-bit i686 cells on
  CircleCI; address and undefined-behaviour sanitizers; a
  build with the pure-C99 multiword-integer fallback forced on
  even where \texttt{\_\_uint128\_t} is available; the Python
  extension on three CPython versions; and an out-of-tree
  \texttt{find\_package} downstream test.  A code-coverage cell
  uploads to Codecov.  A manual build-and-test sweep across all
  current Fedora architectures (\cref{sec:limits}) supplements
  the automated pipeline.
\end{enumerate}

\section{Theory and implementation}
\label{sec:theory}

We follow the conventions of the standard textbook
references~\cite{Edmonds1957__,Varshalovich1988__,Biedenharn1981__}.
All quantum numbers are represented internally and at the
public-API boundary as the integer $2j$ (and similarly $2m$,
$2\ell$). We denote these ``twice-quantum-numbers'' as
$\tilde{j}=2j$, $\tilde{m}=2m$, etc. where notation must
distinguish them from $j$, $m$.

The Wigner 3j, 6j, and 9j symbols, the Clebsch--Gordan
coefficient, and the Racah $W$ coefficient are purely algebraic
SU(2) objects: their values are fixed by the Racah/Wigner
combinatorial formulas, and no spherical-harmonic phase
convention enters the derivation or the implementation. The
Clebsch--Gordan sign convention used here is the
Condon--Shortley convention~\cite{Condon1951__} adopted by
\citet{Edmonds1957__} and \citet{Varshalovich1988__}, given as
\cref{eq:cg} in \cref{sec:cg}, which makes every
Clebsch--Gordan coefficient real. The phase convention for
spherical harmonics enters only the Gaunt coefficient and its
real-spherical-harmonic variant, since both are defined as
integrals of three $Y_\ell^m$. We adopt the standard
Condon--Shortley phase for
$Y_\ell^m$~\cite{Condon1951__,Edmonds1957__,Varshalovich1988__},
and use the explicit Condon--Shortley construction of
\cref{eq:realY-def} for the real spherical harmonics. Users
whose work follows a different real-$Y$ phase convention can
recover the desired result by applying the corresponding
diagonal sign flips at the call site, without modifying the
underlying complex-Gaunt routine.

\subsection{Common building blocks}

Three internal data structures underlie every symbol routine.

\paragraph{Multiword unsigned integers}
\texttt{bigint\_t} is a little-endian array of
\texttt{uint64\_t} words representing an unsigned integer of
arbitrary magnitude. Signed quantities are
stored as a \texttt{bigint\_t} together with a sign flag. The
conversion routines \texttt{bigint\_\allowbreak to\_\allowbreak
float}, \texttt{bigint\_\allowbreak to\_\allowbreak double} and
\texttt{bigint\_\allowbreak to\_\allowbreak long\_double}
extract \texttt{FLT\_MANT\_DIG}, \texttt{DBL\_MANT\_DIG} and
\texttt{LDBL\_MANT\_DIG} bits respectively, with explicit round
and sticky bits, to give correct round-to-nearest-even at the
chosen precision.

\paragraph{Prime-factored rationals}
A \texttt{pfrac\_t} represents a positive rational
$\prod_i p_i^{\exp[i]}$ as the vector of signed integer exponents
$\exp[i]$ indexed by the primes $p_i$ of a precomputed table that
covers every prime up to the largest factorial argument the
workload can reach. The $p$-adic
valuation of $n!$ is computed by Legendre's formula
\begin{equation}
v_p(n!) = \sum_{r=1}^{\infty}\left\lfloor n/p^r\right\rfloor,
\label{eq:legendre}
\end{equation}
so that multiplying a \texttt{pfrac\_t} by $n!$
(\texttt{pfrac\_mul\_factorial}) is a vector addition of
$v_{p_i}(n!)$ to $\exp[i]$ for each $i$, and dividing by $n!$
(\texttt{pfrac\_div\_factorial}) is the corresponding vector
subtraction.  No factorial is ever materialised as an integer.

\paragraph{Exact representation of a symbol}
Every symbol routine produces an internal \texttt{wignernj\_exact\_t}
structure, the seven-tuple
\[
(\,\mathrm{sign},\;\mathrm{sum\_sign},\;\Sigma,\;
N_{\mathrm{int}},\;D_{\mathrm{int}},\;
N_{\mathrm{sqrt}},\;D_{\mathrm{sqrt}}\,),
\]
in which $\Sigma$, $N_{\mathrm{int}}$, $D_{\mathrm{int}}$,
$N_{\mathrm{sqrt}}$, $D_{\mathrm{sqrt}}$ are non-negative
\texttt{bigint\_t}'s and the value of the symbol is
\begin{equation}
\mathrm{sign}\cdot\mathrm{sum\_sign}\cdot
\Sigma\cdot\frac{N_{\mathrm{int}}}{D_{\mathrm{int}}}\cdot
\sqrt{\frac{N_{\mathrm{sqrt}}}{D_{\mathrm{sqrt}}}}.
\label{eq:exact-tuple}
\end{equation}
The conversion to floating point is implemented by a single
\texttt{wignernj\_exact\_to\_double} routine, with analogous
single-precision, long-double, and binary128 variants.  Each
\texttt{bigint\_t} is first reduced to a normalised
mantissa-exponent pair via \texttt{bigint\_frexp}, and the
exponents are combined in integer arithmetic.  The remaining work
is one floating-point square root, two divisions, two
multiplications, and a single \texttt{ldexp}, so the rounding
error is bounded by $O(\varepsilon)$ at the chosen precision. An
analogous \texttt{wignernj\_exact\_to\_mpfr} routine produces an MPFR
result correctly rounded at the user-chosen precision.

\paragraph{Splitting a \texttt{pfrac\_t} into integer and sqrt parts}
Given a \texttt{pfrac\_t} whose value $R$ is the \emph{argument} of an outer
square root (so the symbol carries $\sqrt{R}$), the routine
\texttt{pfrac\_to\_sqrt\_rational} partitions the prime exponents on
parity:
\begin{itemize}
\item each even exponent $\exp[i]=2k_i$ contributes a factor
$p_i^{|k_i|}$ to either $N_{\mathrm{int}}$ (if $k_i>0$) or
$D_{\mathrm{int}}$ (if $k_i<0$);
\item each odd exponent $\exp[i]=2k_i\pm 1$ contributes
$p_i^{|k_i|}$ to $N_{\mathrm{int}}$/$D_{\mathrm{int}}$ as before, and
one additional factor $p_i$ to $N_{\mathrm{sqrt}}$ or
$D_{\mathrm{sqrt}}$.
\end{itemize}
After splitting, $\sqrt{R}=(N_{\mathrm{int}}/D_{\mathrm{int}})\,
\sqrt{N_{\mathrm{sqrt}}/D_{\mathrm{sqrt}}}$ exactly.

\paragraph{Per-thread caches}
Two per-thread caches amortise the setup cost of the symbol
routines across calls.  The first is a multiword scratch backing
the \texttt{pfrac\_t}, least-common-multiple (LCM) exponent
buffers, and \texttt{bigint\_t} workspaces of the Racah sum,
lazily allocated on the first call from a given thread, grown in
place if later calls need wider operands, and reused thereafter.
The second is a per-thread table indexed by $n$ holding the
Legendre exponent vector $\{v_{p_i}(n!)\}_i$, populated lazily as
new $n$ are encountered, so each factorial is decomposed only once
per thread, and subsequent multiplications or divisions by the
same $n!$ reduce to a vector add or subtract over the cached row.
Both caches require thread-local storage (TLS) via one of
\texttt{\_\_thread}, \texttt{\_\_declspec(thread)}, or C11
\texttt{\_Thread\_local}, and toolchains providing none fall back
transparently to allocate-on-call.  An optional
\texttt{wignernj\_warmup\_to} entry point (with its companion
\texttt{wignernj\_thread\_cleanup}) lets callers pre-populate or
relinquish these caches explicitly, and calling either is purely
a performance hint.

\paragraph{Multiword integer kernel}
Multiplication of two \texttt{bigint\_t} operands uses the
schoolbook algorithm below a small word-count threshold and a
\citet{Karatsuba1963_SPD_595} recursion above it. Both share the
same $64\times 64\to 128$ product primitive, taken from the
compiler's \texttt{\_\_uint128\_t} extension where available and
otherwise from a pure-C99 fallback that combines four
$32\times 32\to 64$ partial products with explicit carry tracking
(\cref{sec:limits}).  At typical angular momenta the
operands stay below the Karatsuba crossover, while at large $j$
Karatsuba narrows the gap to the FLINT/GMP (GNU Multiple Precision
Arithmetic Library) back-ends without matching their
sub-quadratic asymptotic.
Division by a 64-bit scalar likewise has two regimes: on x86-64 the
hardware $128/64$ \texttt{divq} instruction is used directly via
inline assembly (the only hardware-specific optimization in the
library), while on every other target the in-tree backend uses
the algorithm of \citet{Moeller2011_ITC_165}, in which a single
precomputed reciprocal is shared across all limbs of one division
so that each per-limb step reduces to two 64-bit multiplications
and a constant number of fixups.  This turns out to be roughly
an order of magnitude cheaper per limb than trial-quotient long
division, and is the reason why the in-tree back-end remains
competitive on non-x86-64 architectures.

A separate fast path covers the case where the dividend is
divisible by the divisor by construction, as occurs in the
small-integer ratio recurrence of Pass~2.  In this regime the
library uses \emph{Hensel exact division}~\cite{Jebelean1993_JSC_169}: a single Newton
iteration mod $2^{64}$ produces $d^{-1}\bmod 2^{64}$, after which
each output limb costs one multiplication by the modular inverse
and one residue subtraction, with no quotient-digit selection or
correction.  Hensel exact division is several times faster per
limb than the non-exact path it replaces, and is the
bottleneck-removing optimisation that makes the Pass~2 ratio
recurrence net-positive at all practical $j$.

\subsection{Wigner 3j symbol}
\label{sec:3j}

The Wigner 3j symbol vanishes unless the projections satisfy
$\tilde{m}_1+\tilde{m}_2+\tilde{m}_3=0$, $|\tilde{m}_i|\le\tilde{j}_i$,
$\tilde{j}_i-\tilde{m}_i$ is even (so $j_i$ and $m_i$ are of the same
half-integer type), and the triangular condition
$|\tilde{j}_1-\tilde{j}_2|\le\tilde{j}_3\le\tilde{j}_1+\tilde{j}_2$
with $\tilde{j}_1+\tilde{j}_2+\tilde{j}_3$ even. Given these
selection rules, the Racah closed-form
expression~\cite{Edmonds1957__,Varshalovich1988__} reads
\begin{equation}
\begin{pmatrix} j_1 & j_2 & j_3 \\ m_1 & m_2 & m_3 \end{pmatrix}
 =
(-1)^{j_1-j_2-m_3}\sqrt{\Delta^2(j_1 j_2 j_3)\,F_{m}}\;S_{3j},
\label{eq:3j-master}
\end{equation}
with the squared triangle coefficient
\begin{equation}
\Delta^2(abc) =
 \frac{(a+b-c)!\,(a-b+c)!\,(-a+b+c)!}{(a+b+c+1)!},
\label{eq:triangle}
\end{equation}
the magnetic-quantum-number factor
\begin{equation}
F_{m} = \prod_{i=1}^{3}(j_i+m_i)!\,(j_i-m_i)!,
\label{eq:Fm}
\end{equation}
and the alternating Racah sum
\begin{equation}
S_{3j} = \sum_{s=s_{\min}}^{s_{\max}}
 \frac{(-1)^s}{s!\,a_1!\,a_2!\,b_1!\,b_2!\,b_3!},
\label{eq:S3j}
\end{equation}
in which
$a_1=j_1+j_2-j_3-s$, $a_2=j_1-m_1-s$, $b_1=j_2+m_2-s$,
$b_2=j_3-j_2+m_1+s$, $b_3=j_3-j_1-m_2+s$, and the summation runs
over all $s$ for which every factorial argument is non-negative.
The triangle condition forces
$\tilde{j}_1+\tilde{j}_2+\tilde{j}_3$ to be even, as is every
$\tilde{j}_i\pm\tilde{m}_i$, so every factorial argument in
\crefrange{eq:triangle}{eq:S3j} is divisible by~2 in the
$\tilde{j}$ representation. The division is performed in integer
arithmetic.
The argument of the outer square root in
\cref{eq:3j-master} is the rational $R\equiv \Delta^2(j_1
j_2 j_3) F_m$.  The routine \texttt{wigner3j\_exact} builds $R$
as a single \texttt{pfrac\_t} via one
\texttt{pfrac\_mul\_factorial} per factor of $\Delta^2 F_m$ and
one \texttt{pfrac\_div\_factorial} for the $(j_1+j_2+j_3+1)!$
denominator.

The Racah sum~\cref{eq:S3j} is evaluated in two passes. For each
term~$s$, the prime-factorization of the term \emph{denominator}
$d_s\equiv s!a_1!a_2!b_1!b_2!b_3!$ is built as a transient \texttt{pfrac\_t} with
non-negative exponents.
\emph{Pass~1} takes, prime-by-prime, the maximum of these
exponents over all $s$,
\begin{equation}
\mathrm{lcm\_exp}[i]=\max_{s}\,v_{p_i}(d_s),
\label{eq:lcm-exp}
\end{equation}
which gives the prime exponents of the LCM
\begin{equation}
L=\prod_i p_i^{\mathrm{lcm\_exp}[i]}.
\label{eq:lcm}
\end{equation}
\emph{Pass~2} converts each rational term $1/d_s$ to the
non-negative integer $L/d_s=\prod_i p_i^{\mathrm{lcm\_exp}[i]-v_{p_i}(d_s)}$, materializes it as a \texttt{bigint\_t}, and accumulates
it with sign $(-1)^s$ into a running pair of unsigned bigint sums
$\Sigma_+$, $\Sigma_-$. To avoid recomputing the prime-power product
from scratch at every $s$, consecutive terms are linked by a
\emph{small-integer ratio recurrence}: the ratio of two consecutive
$|L \cdot \mathrm{term}_s|/|L \cdot \mathrm{term}_{s-1}|$ reduces
to a fixed pattern of small factorial increments and decrements,
so that the ratio collapses to a quotient
$\mathrm{numer}_s/\mathrm{denom}_s$ in which both
$\mathrm{numer}_s$ and $\mathrm{denom}_s$ are products of a small
fixed number of integer factors bounded by $O(j_{\max})$.  The
next \texttt{bigint\_t} is then obtained from its predecessor by
a single batched multiplication by $\mathrm{numer}_s$ followed by
an exact division by $\mathrm{denom}_s$. This
replaces an $O(\pi(j_{\max}))$ prime-power expansion per term with an
$O(1)$ multiply-and-divide. The same recurrence is used in the 6j and
9j Pass-2 evaluators (\cref{sec:6j,sec:9j}) and in the
Gaunt Pass-2 (\cref{sec:gaunt}). The exact division by
$\mathrm{denom}_s$ exploits Hensel-style divisibility-by-construction
arithmetic, described below. The signed difference $\Sigma_+ - \Sigma_-$
produces the final magnitude $\Sigma$ and the sign \texttt{sum\_sign};
the Racah sum is therefore exactly
$\mathrm{sum\_sign}\cdot\Sigma/L$.

The exact tuple of \cref{eq:exact-tuple} is then assembled by
calling \texttt{pfrac\_to\_sqrt\_rational} on $R$ to populate
$N_{\mathrm{int}}$, $D_{\mathrm{int}}$, $N_{\mathrm{sqrt}}$,
$D_{\mathrm{sqrt}}$ and absorbing the LCM denominator $L$ into
$D_{\mathrm{int}}$ by multiplication.  The phase
$\mathrm{sign}=(-1)^{j_1-j_2-m_3}$ is set from the parity of
$(\tilde{j}_1-\tilde{j}_2-\tilde{m}_3)/2$.  The only
floating-point operations in the entire evaluation are the five
\texttt{bigint\_frexp} casts and the four arithmetic operations
plus the single square root and \texttt{ldexp} that combine them
into the final result.

The all-$m$-zero case admits a closed form
\begin{equation}
\begin{pmatrix} j_1 & j_2 & j_3 \\ 0 & 0 & 0 \end{pmatrix}
 = (-1)^g\,\sqrt{\Delta^2(j_1 j_2 j_3)}\,
   \frac{g!}{(g-j_1)!\,(g-j_2)!\,(g-j_3)!},
\label{eq:3j-000-closed}
\end{equation}
with $g=(j_1+j_2+j_3)/2$ (the symbol vanishes when $g$ is
non-integer).  When $\tilde{m}_1=\tilde{m}_2=\tilde{m}_3=0$ the
implementation skips the entire Racah-sum machinery and
materialises the single multinomial-coefficient bigint directly,
which yields a $\sim 22\times$ speedup at $j=5$ growing to
$\sim 190\times$ at $j=1000$ on the benchmark machine of
\cref{sec:benchmarks}.  The Gaunt coefficient inherits the
speedup since one of its two 3j evaluations is always the
$(\ell_1\ell_2\ell_3;0,0,0)$ case.

\subsection{Wigner 6j symbol}
\label{sec:6j}

The 6j symbol $\{\,j_1\,j_2\,j_3;\,j_4\,j_5\,j_6\,\}$ vanishes unless
all four triangles $(j_1 j_2 j_3)$, $(j_1 j_5 j_6)$, $(j_4 j_2 j_6)$,
$(j_4 j_5 j_3)$ are simultaneously satisfied. Its Racah single-sum
representation is
\begin{equation}
\begin{Bmatrix} j_1 & j_2 & j_3 \\ j_4 & j_5 & j_6 \end{Bmatrix}
 = \sqrt{\Delta^2_{1}\Delta^2_{2}\Delta^2_{3}\Delta^2_{4}}\;S_{6j},
\label{eq:6j-master}
\end{equation}
with $\Delta^2_t$ the four squared triangle coefficients of
\cref{eq:triangle} and
\begin{equation}
S_{6j} =
\sum_{s=s_{\min}}^{s_{\max}}\frac{(-1)^s\,(s+1)!}
       {\prod_{t=1}^{4}(s-\alpha_t)!\;\prod_{u=1}^{3}(\beta_u-s)!},
\label{eq:S6j}
\end{equation}
where the four $\alpha$'s are the triangle sums
$j_1+j_2+j_3$, $j_1+j_5+j_6$, $j_4+j_2+j_6$, $j_4+j_5+j_3$, and the
three $\beta$'s are $j_1+j_2+j_4+j_5$, $j_2+j_3+j_5+j_6$,
$j_1+j_3+j_4+j_6$.

The implementation strategy mirrors that of the 3j symbol: the four
$\Delta^2_t$ are accumulated into a single \texttt{pfrac\_t} that represents the
argument $R$ of the outer square root. Because the term in
\cref{eq:S6j} carries the numerator factorial $(s+1)!$ \emph{and}
seven denominator factorials, the prime-factorization of each term
has signed exponents
\begin{equation}
\eta_s[i] \equiv v_{p_i}((s+1)!) - \sum_{t}v_{p_i}((s-\alpha_t)!)
                   - \sum_{u}v_{p_i}((\beta_u-s)!),
\end{equation}
which is the net exponent of $p_i$ in $\mathrm{term}_s$ after
cancellation between the $(s+1)!$ numerator factorial and the
seven denominator factorials---positive when $p_i$ remains in the
numerator and negative when it remains in the denominator.
For each prime $p_i$, the largest power appearing in any term's
denominator is recorded as
\begin{equation}
\mathrm{lcm\_exp}[i] = \max\bigl(0,\,\max_{s}\,(-\eta_s[i])\bigr),
\label{eq:lcm-6j}
\end{equation}
so that the product $L = \prod_i p_i^{\mathrm{lcm\_exp}[i]}$ is
the least common multiple of all term denominators and
$L \cdot \mathrm{term}_s$ is a non-negative integer for every
$s$.  Pass~2 then forms each scaled term as
\begin{equation}
\Bigl|\,L \cdot \mathrm{term}_s\Bigr|
 = \prod_i p_i^{\mathrm{lcm\_exp}[i]+\eta_s[i]} \;\ge\;1,
\end{equation}
which is materialized as a non-negative \texttt{bigint\_t} and
accumulated with sign $(-1)^s$. The exact tuple of
\cref{eq:exact-tuple} is produced exactly as in
\cref{sec:3j}. The overall phase is absorbed into
$(-1)^s$ in the sum so $\mathrm{sign}=+1$.

\subsection{Wigner 9j symbol}
\label{sec:9j}

For the 9j symbol we use the standard reduction to a sum of
products of three 6j symbols,
\begin{align}
\begin{Bmatrix}
j_{11} & j_{12} & j_{13} \\
j_{21} & j_{22} & j_{23} \\
j_{31} & j_{32} & j_{33}
\end{Bmatrix}
&=
(-1)^{j_{13}+j_{22}+j_{31}}\sum_{k}(2k+1)
\begin{Bmatrix} j_{11} & j_{21} & j_{31} \\ j_{32} & j_{33} & k \end{Bmatrix}
\nonumber \\
&\quad\times
\begin{Bmatrix} j_{11} & j_{12} & j_{13} \\ j_{23} & j_{33} & k \end{Bmatrix}
\begin{Bmatrix} j_{22} & j_{21} & j_{23} \\  k     & j_{12} & j_{32} \end{Bmatrix},
\label{eq:9j-as-6j}
\end{align}
with $k$ ranging over values for which all three 6j symbols
satisfy their triangular conditions.  Of the twelve squared
triangle coefficients in the three 6j's, six are independent of
$k$ ($\Delta^2(j_{11}j_{21}j_{31})$, $\Delta^2(j_{32}j_{33}j_{31})$,
$\Delta^2(j_{11}j_{12}j_{13})$, $\Delta^2(j_{23}j_{33}j_{13})$,
$\Delta^2(j_{22}j_{21}j_{23})$, $\Delta^2(j_{22}j_{12}j_{32})$) and
accumulate into a single \texttt{pfrac\_t}~$R$ that becomes the argument of
the \emph{outer} square root.  The remaining three
$k$-dependent $\Delta^2$'s---$\Delta^2(j_{11}j_{33}k)$,
$\Delta^2(j_{21}j_{32}k)$, $\Delta^2(j_{12}j_{23}k)$---each appear
\emph{twice} across the three 6j's (in the two 6j's sharing a
triangle through $k$), so $[\Delta^2]^2$ is a perfect square and
hence rational, and they are absorbed into a per-$k$ \texttt{pfrac\_t}
multiplying the three internal Racah sums.

Concretely, \texttt{wigner9j\_exact} performs:
\begin{enumerate}
\item the six $k$-independent $\Delta^2$ factors are accumulated into
the outer \texttt{pfrac\_t} $R$;
\item over the $k$ loop (\emph{Pass~1}), the per-$k$ \texttt{pfrac\_t}
$P_k=\Delta^2(j_{11}j_{33}k)^2\,\Delta^2(j_{21}j_{32}k)^2\,\Delta^2(j_{12}j_{23}k)^2$
is built (each factor entered twice with
\texttt{add\_delta\_sqrt}, so the resulting exponents are even and
$P_k$ is rational), the three internal 6j Racah sums
$\Sigma^{(1,2,3)}_k$ are computed with their LCMs
$L^{(1,2,3)}_k$, and a \emph{global} LCM exponent vector is built as
\begin{equation}
\mathrm{global\_lcm}[i]=\max_{k}\,\Bigl(L^{(1)}_k[i]+L^{(2)}_k[i]+L^{(3)}_k[i]
                         -\tfrac12\,P_k[i]\Bigr);
\label{eq:9j-glcm}
\end{equation}
\item over the $k$ loop again (\emph{Pass~2}), each contribution
\begin{equation}
(2k+1)\,\Sigma^{(1)}_k\,\Sigma^{(2)}_k\,\Sigma^{(3)}_k\,
\prod_i p_i^{\mathrm{global\_lcm}[i]+\frac12 P_k[i]
            -L^{(1)}_k[i]-L^{(2)}_k[i]-L^{(3)}_k[i]}
\label{eq:9j-term}
\end{equation}
is materialized as a \texttt{bigint\_t} (the prime-power product is
non-negative by construction) and accumulated with sign $\prod_i
\mathrm{sgn}(\Sigma^{(i)}_k)$ into the global signed bigint sum;
\item the final tuple is assembled by calling
\texttt{pfrac\_to\_sqrt\_rational} on $R$ and folding the global LCM
into $D_{\mathrm{int}}$. The overall phase
$(-1)^{j_{13}+j_{22}+j_{31}}$ is set from the parity of
$(\tilde{j}_{13}+\tilde{j}_{22}+\tilde{j}_{31})/2$.
\end{enumerate}
The 9j symbol is therefore delivered in the same exact-tuple form as
the 3j and 6j symbols, and inherits their last-bit accuracy
guarantee.

\subsection{Clebsch--Gordan coefficient}
\label{sec:cg}

The Clebsch--Gordan coefficient is computed from the 3j symbol as
\begin{equation}
\langle j_1 m_1\,j_2 m_2 | J M\rangle
 = (-1)^{j_1-j_2+M}\sqrt{2J+1}
\begin{pmatrix} j_1 & j_2 & J \\ m_1 & m_2 & -M \end{pmatrix}.
\label{eq:cg}
\end{equation}
Internally, \texttt{clebsch\_gordan\_exact} first invokes the 3j
exact-pipeline of \cref{sec:3j}, then folds the
$\sqrt{2J+1}=\sqrt{\tilde{J}+1}$ factor into the resulting tuple
\emph{before} the floating-point cast: the integer $\tilde{J}+1$ is
trial-divided by the primes in turn, and for each prime power
$p^{c}$ in $\tilde{J}+1$ the contribution $p^{\lfloor c/2\rfloor}$
is multiplied into $N_{\mathrm{int}}$ and, if $c$ is odd, an extra
factor $p$ is multiplied into $N_{\mathrm{sqrt}}$. The
Clebsch--Gordan coefficient therefore inherits the same last-bit
accuracy as the underlying 3j symbol, since the only added work is
exact integer factorization.

\subsection{Racah $W$ coefficient}
\label{sec:racah}

The Racah $W$ coefficient is the original (sign-different) form of
the 6j symbol,
\begin{equation}
W(j_1\,j_2\,J\,j_3;\,j_{12}\,j_{23})
 = (-1)^{j_1+j_2+j_3+J}
 \begin{Bmatrix} j_1 & j_2 & j_{12} \\ j_3 & J & j_{23} \end{Bmatrix},
\label{eq:racah}
\end{equation}
and is implemented as a thin wrapper around the 6j routine.

\subsection{Fano $X$-coefficient}
\label{sec:fanox}

The Fano $X$-coefficient~\cite{Fano1951__,Fano1959__}
is a normalisation variant of the 9j symbol introduced for the
analysis of polarisation correlations:
\begin{equation}
X(j_1\,j_2\,j_{12};\,j_3\,j_4\,j_{34};\,j_{13}\,j_{24}\,J)
 = \sqrt{(2j_{12}+1)(2j_{34}+1)(2j_{13}+1)(2j_{24}+1)}
   \begin{Bmatrix}
     j_1 & j_2 & j_{12} \\
     j_3 & j_4 & j_{34} \\
     j_{13} & j_{24} & J
   \end{Bmatrix}.
\label{eq:fanox}
\end{equation}
The four $\sqrt{2j+1}$ factors are folded into the existing 9j
exact tuple in the same way the single $\sqrt{2J+1}$ factor of the
Clebsch--Gordan coefficient is folded into the 3j tuple
(\cref{sec:cg}): each $(2j+1)$ is decomposed into prime
powers and distributed across $N_{\mathrm{int}}$ and
$N_{\mathrm{sqrt}}$ on parity. The selection rules and asymptotic
cost are those of the underlying 9j symbol.

\subsection{Gaunt coefficient}
\label{sec:gaunt}

The Gaunt coefficient over complex spherical harmonics,
\begin{equation}
\mathcal{G}^{m_1 m_2 m_3}_{\ell_1\ell_2\ell_3}
 \equiv \int Y_{\ell_1}^{m_1}(\Omega) Y_{\ell_2}^{m_2}(\Omega)
            Y_{\ell_3}^{m_3}(\Omega)\,\mathrm{d}\Omega,
\label{eq:gaunt-def}
\end{equation}
is conventionally written in terms of two 3j symbols,
\begin{equation}
\mathcal{G}^{m_1 m_2 m_3}_{\ell_1\ell_2\ell_3}
 = \sqrt{\frac{(2\ell_1+1)(2\ell_2+1)(2\ell_3+1)}{4\pi}}
   \begin{pmatrix}\ell_1&\ell_2&\ell_3\\ 0&0&0\end{pmatrix}
   \begin{pmatrix}\ell_1&\ell_2&\ell_3\\ m_1&m_2&m_3\end{pmatrix}.
\label{eq:gaunt-3j}
\end{equation}
Calling the 3j routine twice would give correctly-rounded results,
but a tighter pipeline combines everything into a single \texttt{pfrac\_t}.
Substituting \cref{eq:3j-master} into
\cref{eq:gaunt-3j}, the two outer square roots multiply
together so that $\Delta^2(\ell_1\ell_2\ell_3)$ appears
\emph{twice}---i.e.\ as the rational $[\Delta^2]^2$---under the
combined outer square root, and the magnetic factor $F_{m=0} =
\prod_i (\ell_i!)^2$ from the $(0,0,0)$ symbol is likewise a pure
rational.  The combined outer square root in the Gaunt evaluation
has argument
\begin{equation}
[\Delta^2(\ell_1\ell_2\ell_3)]^2\;
\prod_{i=1}^{3}(\ell_i!)^2\;
\prod_{i=1}^{3}(\ell_i+m_i)!(\ell_i-m_i)!\;
\frac{(2\ell_1+1)(2\ell_2+1)(2\ell_3+1)}{4},
\label{eq:gaunt-outer}
\end{equation}
which is a pure rational. The factor $1/\sqrt{\pi}$ from the spherical
harmonic normalization is the only irrational quantity in the entire
expression. It is applied as a single multiplication at the
floating-point step, after the bigint-to-float cast.

The implementation \texttt{gaunt\_exact} therefore proceeds as
follows.  The combined outer \texttt{pfrac\_t} is built by adding the squared
triangle coefficient \emph{twice} (so its prime exponents are all
even and it folds entirely into
$N_{\mathrm{int}}/D_{\mathrm{int}}$ at the sqrt-rational split),
then the six factorials of $F_{m=0}$ and the six of $F_m$, the
three integer factors $(2\ell_i+1)$, and the integer $1/4$
(realised by subtracting~2 from the exponent of $p_1=2$).  Since
the squared triangle coefficient and all magnetic-quantum-number
factorials are already in the outer pfrac, the integer cores of
the two 3j symbols reduce to two independent sums, computed with
their own LCM denominators: the $(\ell_1\ell_2\ell_3;0,0,0)$ core
collapses to the closed-form multinomial coefficient
$g!/[(g-\ell_1)!(g-\ell_2)!(g-\ell_3)!]$ from
\cref{eq:3j-000-closed} (with $g=(\ell_1+\ell_2+\ell_3)/2$),
which the implementation materialises directly without entering
the Racah loop, while the $(\ell_1\ell_2\ell_3;m_1,m_2,m_3)$ core
is the alternating Racah single sum $S_{3j}$ of \cref{eq:S3j}.
The two bigint magnitudes are multiplied together to give
$\Sigma$, and the LCM exponents are summed and folded into
$D_{\mathrm{int}}$.  The phase reduces, after
simplification of the two 3j phases, to
$(-1)^{m_3}=(-1)^{\tilde{m}_3/2}$.  At the final floating-point
step the result is divided by $\sqrt{\pi}$ computed in the target
precision.

\subsection{Gaunt coefficient over real spherical harmonics}
\label{sec:gaunt-real}

Quantum-chemical and electromagnetic codes often work with real
spherical harmonics rather than complex ones. We adopt the
Condon--Shortley / Wikipedia convention,
\begin{align}
S_{\ell,0}     &= Y_\ell^{0}, \nonumber\\
S_{\ell,m>0}   &= \tfrac{1}{\sqrt 2}\bigl(Y_\ell^{-m} + (-1)^m Y_\ell^{m}\bigr),
   \label{eq:realY-def}\\
S_{\ell,m<0}   &= \tfrac{i}{\sqrt 2}\bigl(Y_\ell^{m} - (-1)^{|m|} Y_\ell^{-m}\bigr),
\nonumber
\end{align}
and define the real Gaunt coefficient as the corresponding triple
integral $\mathcal{G}^{R\,m_1 m_2 m_3}_{\ell_1\ell_2\ell_3}\equiv\int
S_{\ell_1 m_1} S_{\ell_2 m_2} S_{\ell_3 m_3}\,\mathrm{d}\Omega$. Substituting
\Cref{eq:realY-def} expresses $\mathcal{G}^R$ as a linear
combination of complex Gaunts at the same $(\ell_1,\ell_2,\ell_3)$ but
with $m_i \to s_i\lvert m_i\rvert$ for sign assignments
$s_i\in\{\pm1\}$ that satisfy
$\sum_i s_i\lvert m_i\rvert = 0$ \cite{Homeier1996_JMST_31}. Up to
eight sign tuples are scanned and at most two contribute. Each
contributing tuple carries a coefficient of the form
$\tau\,\varphi/\sqrt{2}^{\,k}$, with $\tau\in\{\pm 1\}$,
$\varphi\in\{1,i\}$, and $k$ equal to the number of non-zero
$\lvert m_i\rvert$.

Two reductions make this run at the cost of \emph{one} complex-Gaunt
evaluation. (i) The valid sign tuples form a sign-flipped pair
$(\mathbf{s},-\mathbf{s})$, and the complex Gaunt is invariant
under simultaneous $m_i$ flip when $\ell_1+\ell_2+\ell_3$ is
even---which is required for any non-vanishing Gaunt---so both
tuples evaluate to the same $\mathcal{G}^C$ and their coefficients
can be summed before calling the complex routine. (ii) The
combined prefactor is rational or $\sqrt{2}$ times a rational with
small numerator and denominator, absorbed exactly into
\texttt{wignernj\_exact\_t} by multiplying $N_{\mathrm{int}}$,
$D_{\mathrm{int}}$, and $D_{\mathrm{sqrt}}$ by small integers.
Last-bit accuracy is therefore preserved for the real Gaunt
without any additional floating-point work beyond the single
$1/\sqrt{\pi}$ multiplication of the complex pipeline.  If no
sign tuple satisfies the constraints, or the two coefficients
cancel, \texttt{gaunt\_real} returns $0$ without entering the
Racah sum.

\subsection{libquadmath and MPFR interfaces}
\label{sec:mpfr}

The exact intermediate representation of every symbol given in
\cref{eq:exact-tuple} is independent of the target floating-point
precision, so additional precisions plug in by swapping only the
final cast.  \texttt{libwignernj} provides an optional libquadmath
back-end (enabled with \texttt{-DWIGNERNJ\_BUILD\_QUADMATH=ON}) exposing
IEEE~754 binary128 (\texttt{\_\_float128}, 113-bit mantissa,
$\sim 16400$ minimum decimal exponent) on toolchains that ship
\texttt{<quadmath.h>}. The conversion path
\texttt{wignernj\_exact\_to\_float128} mirrors the long-double
variant, with \texttt{bigint\_to\_float128} assembling the top
three 64-bit words of the bigint via Horner-style binary128
arithmetic, feeding 192 input bits into a 113-bit mantissa and
leaving $\sim 79$ bits of headroom against pathological
tied-half-ulp cases (Apple Clang and MSVC lack
\texttt{\_\_float128} and the back-end is silently omitted there).
The optional MPFR interface~\cite{Fousse2007_ATMS_13} takes the
same exact tuple and casts it through \texttt{bigint\_to\_mpfr}
and the MPFR arithmetic primitives at the user-specified precision
and rounding mode. The MPFR Gaunt routine additionally evaluates
$\sqrt{\pi}$ via
\texttt{mpfr\_const\_pi}/\texttt{mpfr\_sqrt}.

\subsection{Implementation limits and portability}
\label{sec:limits}

Two limits constrain the angular momenta accessible to
\texttt{libwignernj}.
The first is the \emph{prime sieve}.  Every factorial that can
appear in any symbol must factor entirely into primes that are
present in a precomputed table.  We hard-code that table into the
compiled library rather than building it at run time: this
realises the no-caller-side-initialisation aim, makes the library
safely usable from concurrent threads with no global-init race,
and lets the table live in read-only data with no allocator
activity on first use.  The default sieve limit is
\texttt{PRIME\_SIEVE\_LIMIT}~$=20011$ (2263 primes), chosen so the
prime list and the inverse-lookup index together fit in a
rule-of-thumb 50~kB compile-time-constant budget.  The largest
factorial argument that can then be safely formed is
\texttt{MAX\_FACTORIAL\_ARG}~$=20020$.  In angular-momentum terms
the largest factorial encountered in a 3j, 6j, Clebsch--Gordan,
Racah $W$, complex Gaunt, or real-spherical-harmonic Gaunt symbol
is $(j_1+j_2+j_3+1)!$, giving the default-build ceiling $j_1 + j_2
+ j_3 \le 20019$ (or $j\le 6673$ at equal $j$). For 9j the
$k$-dependent triangle denominators reach $(4j+1)!$ in the
equal-$j$ limit, giving $j\le 5004$.  Calls beyond these bounds
abort with a diagnostic message rather than silently returning a
wrong result.  These ranges already cover the angular-momentum
needs of the application domains in the introduction. When larger
angular momenta are required, the ceiling can be raised by
regenerating the prime table with a larger sieve limit via
\texttt{tools/gen\_prime\_table.py} and rebuilding.

The second limit is computational.  The Racah sum has $O(j)$
terms for the 3j and 6j, and the 9j adds an outer $k$ loop of
length $O(j)$, so the term count scales as $O(j)$ for 3j and 6j,
and $O(j^2)$ for 9j.  The intermediate multiword integers grow
linearly in $j$ ($\sim j/\ln 2$ bits per bigint, since the LCM
denominator grows as the primorial), so each elementary bigint
operation already costs $O(j)$.  Combining the two factors, the
asymptotic per-symbol cost is $O(j^2)$ for 3j, 6j, Clebsch--Gordan,
Racah $W$, and the complex and real-Gaunt routines, and $O(j^4)$
for 9j.  On modern hardware the practical horizons are: 3j and 6j
in milliseconds up to $j\sim 1000$ and seconds up to $j\sim 6000$;
9j in milliseconds up to $j\sim 100$ and minutes up to
$j\sim 1000$.

Regarding portability, the multiword integer routines internally
require a $64\times 64 \to 128$ multiplication and a $128/64$
division primitive.  When the compiler advertises
\texttt{\_\_SIZEOF\_INT128\_\_} these are obtained directly from
the \texttt{\_\_uint128\_t} extension. Otherwise (for instance, in the case of MSVC) a
pure-C99 fallback is selected
automatically, decomposing the $64\times 64$ product into four
$32\times 32$ partial products with explicit carry tracking and
implementing the $128/64$ division (whose divisor here is always a
prime $\le\,$\texttt{PRIME\_SIEVE\_LIMIT} and so fits in 32 bits)
as two $64/32$ long-division steps.  Both paths produce
bit-identical output.

The library is also \emph{endianness-agnostic}.  All arithmetic is
performed at the value level through standard C operators on
\texttt{uint64\_t}. No code casts an integer pointer to
\texttt{uint8\_t*} or accesses individual bytes of a multiword
integer.  The ``little-endian'' descriptor in
\texttt{src/bigint.h} refers to logical word ordering
(\texttt{words[0]} is the least-significant base-$2^{64}$ digit),
not to the host CPU's byte order.  As a concrete cross-architecture
verification, the package has been manually built and the entire
test suite run to completion on all current Fedora architectures:
the little-endian \texttt{i686}, \texttt{x86\_64}, \texttt{aarch64},
and \texttt{ppc64le} targets and the big-endian \texttt{s390x}
target.

\section{Software organization and language interfaces}
\label{sec:software}

The library is organised as a small C core
\texttt{libwignernj} with thin wrappers that expose the same
functions in C++, Fortran 90, and Python.  Every public function
takes the integer arguments $\tilde{j} = 2j$ (and analogously
$\tilde{m}$, $\tilde{\ell}$).  A vanishing symbol returns $0$, so a
selection-rule violation is not an error.  The build system is
CMake, and the only mandatory build dependency is a C99 compiler;
optional dependencies are picked up via
\texttt{find\_package} when their respective options are enabled.

\subsection{C API}

For every symbol three functions are provided---one per IEEE 754
binary precision---with the precision encoded in the suffix:
\texttt{\_f} for single, \texttt{\_l} for long double, and no
suffix for double. The optional libquadmath and MPFR back-ends
described below add \texttt{\_q} and \texttt{\_mpfr} variants.
The header \texttt{wignernj.h} declares
\begin{lstlisting}[language=C]
double wigner3j  (int tj1, int tj2, int tj3,
                  int tm1, int tm2, int tm3);
double wigner6j  (int tj1, int tj2, int tj3,
                  int tj4, int tj5, int tj6);
double wigner9j  (int tj11, int tj12, int tj13,
                  int tj21, int tj22, int tj23,
                  int tj31, int tj32, int tj33);
double clebsch_gordan(int tj1, int tm1, int tj2, int tm2,
                       int tJ,  int tM);
double racah_w   (int tj1, int tj2, int tJ, int tj3,
                  int tj12, int tj23);
double fano_x    (int tj1, int tj2, int tj12,
                  int tj3, int tj4, int tj34,
                  int tj13, int tj24, int tJ);
double gaunt     (int tl1, int tm1, int tl2, int tm2,
                  int tl3, int tm3);
double gaunt_real(int tl1, int tm1, int tl2, int tm2,
                  int tl3, int tm3);
\end{lstlisting}
together with the \texttt{\_f} and \texttt{\_l} variants. A
\texttt{pkg-config} file \texttt{libwignernj.pc} is installed
alongside the library so that downstream projects can locate the
include and link flags through
\texttt{pkg-config --cflags --libs libwignernj}.

The header also exposes a small set of optional cache-management
entry points for callers who want explicit control over the
per-thread caches described in \cref{sec:theory}.
\texttt{wignernj\_warmup\_to(N)} pre-grows both per-thread caches
(the Racah-pipeline scratch and the factorial-decomposition cache)
to fit any subsequent symbol evaluation whose worst-case factorial
argument is bounded by $N$.  Passing $N = 0$ sizes the caches to
the absolute prime-table ceiling.  \texttt{wignernj\_thread\_cleanup()}
releases both caches for the calling thread when it is finished
computing symbols.  The eight companion helpers
\texttt{wigner3j\_max\_factorial}, \dots,
\texttt{gaunt\_real\_max\_factorial} return the largest factorial
argument that the corresponding symbol would reference for the
given inputs, so that the warmup can be sized exactly for the
workload at hand. None of these calls is required: when omitted,
both caches lazy-grow on first use.

\subsection{Optional libquadmath back-end}

When the library is built with \texttt{-DWIGNERNJ\_BUILD\_QUADMATH=ON}, an
auxiliary header \texttt{wignernj\_quadmath.h} is installed and a
\texttt{\_\_float128} variant of every public symbol
(\texttt{wigner3j\_q}, \dots, \texttt{gaunt\_real\_q}) is exposed.
The IEEE~754 binary128 type is a fixed-precision compiler builtin,
so unlike the MPFR back-end of \cref{sec:mpfr-api} the caller
does not need to allocate or destroy a result object:
\begin{lstlisting}[language=C]
#include "wignernj_quadmath.h"
__float128 v = wigner6j_q(4, 4, 4, 4, 4, 4);
\end{lstlisting}
CMake detects \texttt{<quadmath.h>} and links libquadmath
automatically. On toolchains without \texttt{\_\_float128} support
(e.g.\ Apple Clang, MSVC) the option is rejected with a clear
error.
The native binary128 type is also propagated to the Fortran
interface as \texttt{real(c\_float128)} from
\texttt{iso\_c\_binding}, the gfortran/Intel~ifx extension that
maps directly to C's \texttt{\_\_float128}.  This is the same
physical kind as \texttt{real128} from \texttt{iso\_fortran\_env},
but \texttt{c\_float128} is the formally C-interoperable name and
avoids a \texttt{-Wc-binding-type} warning on gfortran~16.  The
\texttt{wignernj} module gains the corresponding
\texttt{wigner3j\_q}, \dots, \texttt{gaunt\_real\_q}
\texttt{bind(c)} interfaces and the real-valued convenience
wrappers \texttt{w3jq}, \dots, \texttt{wgaunt\_realq}, all
available through the same \texttt{use wignernj} that exposes
the double-precision routines.

\subsection{Optional MPFR back-end}
\label{sec:mpfr-api}

When the library is built with \texttt{-DWIGNERNJ\_BUILD\_MPFR=ON}, an
additional header \texttt{wignernj\_mpfr.h} is installed and an
\texttt{mpfr\_t} variant of every public symbol
(\texttt{wigner3j\_mpfr}, \dots, \texttt{gaunt\_real\_mpfr}) is
exposed.  The caller sets the desired precision on the output
\texttt{mpfr\_t} via \texttt{mpfr\_init2} before calling the
function. The rounding mode is supplied as the last argument and
may be any of the standard MPFR modes
\texttt{MPFR\_RNDN}, \texttt{MPFR\_RNDZ}, \texttt{MPFR\_RNDD},
\texttt{MPFR\_RNDU}, or \texttt{MPFR\_RNDA}, with
round-to-nearest the usual choice for numerical work:
\begin{lstlisting}[language=C]
mpfr_t v;
mpfr_init2(v, 256);                 /* 256-bit precision */
wigner6j_mpfr(v, 4, 4, 4, 4, 4, 4,  MPFR_RNDN); /* round to nearest, ties to even */
mpfr_clear(v);
\end{lstlisting}
The MPFR back-end is the recommended way to obtain results at
precisions higher than the platform's \texttt{long double} when an
extreme angular momentum exposes the finite-mantissa-width error
of the floating-point cast.  The interface is intentionally a
C-API feature only.  The C\texttt{++} and Fortran~90 bindings of
\cref{sec:software} expose \texttt{float}, \texttt{double}, and
\texttt{long double}; the Python binding exposes only
\texttt{float} and \texttt{double}, since CPython's built-in
\texttt{float} type is a C \texttt{double} and there is no native
container for anything wider (\cref{sec:python}).  No canonical
arbitrary-precision binding appears to exist in any of the three
languages to which the wrappers could couple without forcing a
dependency on every downstream user.
Consumers in those languages who need higher-than-double precision
should call the C \texttt{*\_mpfr} (or \texttt{*\_q}) routines
directly through the foreign-function-interface mechanism of their
environment (e.g.\ \texttt{ctypes} or \texttt{cffi} from Python,
\texttt{iso\_c\_binding} from Fortran, or a plain \texttt{extern
"C"} declaration from C++).

\subsection{C++ wrapper}

A header-only C++11 wrapper \texttt{wignernj.hpp} provides a typed,
overloaded interface that links against the C library
(\texttt{-lwignernj -lm}). The wrapper has no separate translation
unit: each templated function calls the underlying C symbol of the
appropriate precision, so adding it to a project costs at most an
\texttt{\#include}. Two calling conventions are accepted:
\begin{lstlisting}[language=C++]
// Integer 2*j convention (matches the C API)
double v = wignernj::symbol3j<double>(2, 2, 0,  0, 0, 0);
float  f = wignernj::symbol6j<float>(2, 2, 2,  2, 2, 2);

// Real-valued convention (must be exact half-integers)
double v = wignernj::symbol3j(1.0, 1.0, 0.0,  0.0, 0.0, 0.0);
double c = wignernj::cg(0.5, 0.5,  0.5, -0.5,  1.0, 0.0);
\end{lstlisting}
The real-valued overloads check that every floating-point
argument is an exact half-integer and throw a
\texttt{std::}\allowbreak\texttt{invalid\_argument} otherwise.  The available
functions span the eight symbol families
(\texttt{symbol3j}, \dots, \texttt{gauntreal}).

\subsection{Fortran 90 module}

The Fortran interface uses the \texttt{iso\_c\_binding} module, so
every symbol is a direct \texttt{bind(c)} interface to the C
library.  The lower layer mirrors the C API at the
$\tilde{j}$-integer level for all standard precisions (plus the
optional \texttt{\_q} variant when libquadmath is enabled). An
upper layer of real-valued wrappers \texttt{w3j}, \dots,
\texttt{wgaunt\_real} takes \texttt{real(8)} arguments and
convert to $\tilde{j}$ internally:
\begin{lstlisting}[language=Fortran]
use wignernj
real(8) :: v
v = w3j(1.0d0, 1.0d0, 0.0d0,  0.0d0, 0.0d0, 0.0d0)
v = w6j(1.0d0, 1.0d0, 2.0d0,  1.0d0, 1.0d0, 2.0d0)
v = wcg(0.5d0, 0.5d0,  0.5d0, -0.5d0,  1.0d0, 0.0d0)
v = wgaunt(2.0d0, 1.0d0, 2.0d0, -1.0d0, 2.0d0, 0.0d0)
v = wgaunt_real(2.0d0, 1.0d0, 2.0d0, -1.0d0, 0.0d0, 0.0d0)
\end{lstlisting}
The Fortran interface is built into a separate library
\texttt{libwignernj\_f03} so that pure-C consumers do not have to
link against a Fortran runtime.  When the C library is built with
\texttt{-DWIGNERNJ\_BUILD\_QUADMATH=ON}, the module additionally exposes
\texttt{wigner3j\_q}, \dots, \texttt{gaunt\_real\_q}, returning
\texttt{real(c\_float128)} from \texttt{iso\_c\_binding}.
Companion real-valued wrappers \texttt{w3jq}, \dots,
\texttt{wgaunt\_realq} accept the same type and call the
\texttt{bind(c)} interfaces internally.  On toolchains without
binary128 those declarations are preprocessed out and the module
compiles with only the
\texttt{float}/\texttt{double}/\texttt{long~double} layer.

\subsection{Python extension}
\label{sec:python}

A CPython extension module is provided as a self-contained source
distribution that builds and installs through standard
\texttt{pip install -e~.} (the Python library can also be built
with CMake).
The extension accepts integer, floating-point, and
\texttt{fractions.Fraction} arguments interchangeably and dispatches
to the appropriate C precision through the \texttt{precision=}
keyword:
\begin{lstlisting}[language=Python]
import wignernj
from fractions import Fraction
wignernj.wigner3j(1, 1, 0,  0, 0, 0)
wignernj.wigner3j(0.5, 0.5, 1,  0.5, -0.5, 0)
wignernj.wigner6j(1, 1, 2,  1, 1, 2)
wignernj.wigner9j(1, 1, 2,  1, 1, 2,  2, 2, 4)
wignernj.clebsch_gordan(1, 1,  1, -1,  2, 0)
wignernj.racah_w(1, 1, 0, 1,  0, 1)
wignernj.fano_x(1, 1, 1,  1, 1, 1,  1, 1, 1)
wignernj.gaunt(2, 1, 2, -1, 2, 0,  precision='float')
wignernj.gaunt_real(2, 1, 2, -1, 0, 0)
wignernj.wigner3j(Fraction(1, 2), Fraction(1, 2), 1,
                Fraction(1, 2), Fraction(-1, 2), 0)
\end{lstlisting}
The Python wrapper validates that every floating-point or
\texttt{Fraction} argument represents an exact half-integer, raising
\texttt{ValueError} otherwise.

The Python extension recompiles the C99 sources directly into
\texttt{\_wignernj.so} instead of linking against
\texttt{lib\allowbreak wignernj}, producing a single
self-contained shared object.  The pip wheel built from \texttt{setup.py} is the
canonical distribution path and deliberately uses the in-tree
bigint with no optional dependencies, so \texttt{pip install}
works out of the box on every supported platform.  The
\texttt{precision=} keyword accepts only \texttt{'float'} and
\texttt{'double'}: CPython's built-in \texttt{float} type is a
C \texttt{double}, and no native CPython container exists for
binary128 or arbitrary precision, so a \texttt{'longdouble'}
or \texttt{'quad'} option would have to box through
\texttt{PyFloat\_FromDouble} and silently truncate to double.
Callers who need higher-than-double precision from Python
should call the C \texttt{*\_q} (libquadmath, IEEE~754
binary128) or \texttt{*\_mpfr} (arbitrary precision) routines
directly via a foreign-function-interface mechanism such as
\texttt{ctypes} or \texttt{cffi}, paired with a Python
binding for the chosen precision (\texttt{gmpy2} or
\texttt{mpmath} for MPFR); we have not bundled such a binding
because no single canonical multi-precision binding exists
across the Python ecosystem.
The CMake-driven Python build
(\texttt{cmake -DWIGNERNJ\_BUILD\_PYTHON=ON -DWIGNERNJ\_BUILD\_FLINT=ON}), in
contrast, can compile FLINT support into \texttt{\_wignernj.so},
giving the Python module the same sub-quadratic asymptotic at
large angular momenta as the C library, at the cost of a runtime
dependency on FLINT, GMP, and GNU~MPFR.

\subsection{Build system and licensing}

The CMake build (out-of-tree:
\texttt{cmake -B build \&\& cmake --build build}) honours the
following options:
\begin{description}
\item[\texttt{WIGNERNJ\_\allowbreak BUILD\_COVERAGE}] build with \texttt{--coverage -O0}
  for \texttt{lcov} / Codecov instrumentation.
\item[\texttt{WIGNERNJ\_\allowbreak BUILD\_CXX\_TESTS}] build and run the C++
  wrapper-test programs.
\item[\texttt{WIGNERNJ\_\allowbreak BUILD\_EXAMPLES}] build and register the per-binding
  example programs as ctest tests (default on).
\item[\texttt{WIGNERNJ\_\allowbreak BUILD\_FLINT}] enable the optional FLINT bigint
  back-end~\cite{Hart2010__88}: sub-quadratic multiplication via
  GMP's Karatsuba / Toom--Cook / Sch\"onhage--Strassen ladder;
  pulls in FLINT along with its GMP and GNU~MPFR transitive
  dependencies.
\item[\texttt{WIGNERNJ\_\allowbreak BUILD\_FORTRAN}] build the Fortran 90 module
  (\texttt{libwignernj\_f03}).
\item[\texttt{WIGNERNJ\_\allowbreak BUILD\_LTO}] enable
  inter-procedural (IPO) / link-time (LTO) optimisation; probes
  the toolchain for support and silently disables it if it is
  unavailable.
\item[\texttt{WIGNERNJ\_\allowbreak BUILD\_MPFR}] build the optional arbitrary-precision
  back-end (pulls in GNU~MPFR~\cite{Fousse2007_ATMS_13}).
\item[\texttt{WIGNERNJ\_\allowbreak BUILD\_PYTHON}] build the
  CPython extension via CMake; needs \texttt{setuptools} at
  build time.  An alternative to the \texttt{pip install} path.
\item[\texttt{WIGNERNJ\_\allowbreak BUILD\_QUADMATH}] build the optional libquadmath
  binary128 back-end.
\item[\texttt{BUILD\_SHARED\_LIBS}] build a shared library
  (default); set \texttt{OFF} for a static archive.
\item[\texttt{WIGNERNJ\_\allowbreak BUILD\_TESTS}] build and run the C and Fortran test
  suites.
\end{description}

The library is released under the BSD~3-Clause licence and has no
external runtime dependencies in its default configuration. The
optional back-ends pull in their listed dependencies as noted
above.  The codebase is small and compiles quickly: the C99 core,
the C++ header-only wrapper, and the Fortran 90 module together
amount to roughly four thousand lines of source code. A clean
parallel build of the library plus all four language bindings
completes in $\sim 1\,$s on a recent x86-64 laptop, and the
resulting shared library is approximately 125~kB on x86-64 Linux,
of which roughly half is the compiled-in prime table.  A separate
preprocessor switch \texttt{-DBIGINT\_FORCE\_PORTABLE} forces the
multiword-integer back-end onto its pure-C99 fallback even on
compilers that support \texttt{\_\_uint128\_t}. This flag is
exercised in CI to verify bit-identical output across the two
paths.

Several auxiliary directories support reproducibility and
downstream consumption.  The \texttt{benchmarks/} directory
contains library-versioning microbenchmarks and profile
drivers used in the development of \texttt{libwignernj}
itself.  The comparative bench
that produced \cref{fig:accuracy,fig:timing} links against
\texttt{WIGXJPF} and GSL, neither of which is a dependency of
\texttt{libwignernj}.  We therefore distribute it as
supplementary material to this work rather than in the main
repository.  \texttt{tests/\allowbreak cmake\_\allowbreak
downstream/} is a minimal out-of-tree downstream project (one
C, one C++, and one Fortran source plus a six-line
\texttt{CMakeLists.txt}) that consumes \texttt{libwignernj}
through CMake's \texttt{find\_\allowbreak package} mechanism
with the \texttt{Fortran} component requested, and is exercised
in CI on every push.
\texttt{examples/} ships a single-file demonstration of every
public symbol family in each of the four bindings, registered as a
ctest test when \texttt{WIGNERNJ\_BUILD\_EXAMPLES=ON} (the default).
\texttt{tests/gen\_refs.py} regenerates the
\texttt{sympy}-reference tables of
\cref{sec:verification}, and \texttt{tools/} contains the
prime-table and source-list generators used at build time.

\section{Verification and testing}
\label{sec:verification}

Verification of an exact-arithmetic library has two distinct
aims: that the returned values are correct and that the
floating-point conversion is correctly rounded.
\texttt{libwignernj} addresses both through the multi-pronged
\texttt{tests/} suite, which ships with the source and is
exercised on every push and pull request.

\paragraph{Reference values from \texttt{sympy}}
The first level of verification is a reference table of a few
thousand symbols per family (3j, 6j, 9j, Clebsch--Gordan, Racah
$W$, complex Gaunt, real-spherical-harmonic Gaunt), generated
offline from
\texttt{sympy.\allowbreak physics.\allowbreak wigner}~\cite{Meurer2017_PCS_103}
using exact rational arithmetic and stored as C source files
(\texttt{tests/test\_3j.c}, etc.).  Each test compares the
library's value against the reference. Because the internal
arithmetic is exact, the only divergence is the final rounding, so
the tolerance is in practice a few units in the last place (ULPs).
The Fano $X$-coefficient is not in \texttt{sympy} and is verified
instead by a self-consistency oracle
($X = \sqrt{(2j_{12}+1)(2j_{34}+1)(2j_{13}+1)(2j_{24}+1)}\cdot
\{9j\}$ on a sweep of valid inputs) in
\texttt{tests/test\_derived.c}.  The reference table is
regenerated by \texttt{tests/gen\_refs.py}.

\paragraph{Symmetry oracles}
A complementary test (\texttt{tests/test\_symmetry.c}) checks the
well-known permutation/phase symmetries of each symbol on a
deterministic-seed pseudorandom sweep of valid configurations. The
3j symbol must satisfy cyclic invariance under column rotation, a
$(-1)^J$ phase under odd column swap and under the simultaneous
sign-flip of all $m$. The 6j satisfies column-permutation symmetry;
and the 9j is symmetric under any row or column permutation up to
a $(-1)^{\Sigma j}$ phase for an odd permutation. These identities
each combine many independent intermediate quantities, so they are
sensitive to sign-bit errors, factor-of-$(-1)^J$ phase mistakes, and
Racah-sum-bound asymmetries that a hand-picked reference table can
miss.

\paragraph{Multi-precision agreement}
\texttt{tests/test\_precisions.c} verifies that the \texttt{float},
\texttt{double}, and \texttt{long~double} variants agree within
their respective precisions and round to the same value when
reduced to the lowest. The binary128 variants are exercised in
\texttt{tests/test\_quadmath.c}, and the MPFR back-end in
\texttt{tests/test\_mpfr.c}, where the 64-bit result is compared
to the C \texttt{double} output and the 256-bit result to a
closed-form analytic value within two ULP.

\paragraph{Building blocks}
The internal modules \texttt{bigint.c}, \texttt{pfrac.c}, and
\texttt{primes.c} have dedicated tests
(\texttt{test\_bigint.c}, \texttt{test\_pfrac.c},
\texttt{test\_primes.c}) that cover the full bigint arithmetic
surface (with explicit IEEE~754 round/sticky-bit verification on
the mantissa-extraction routines), check that
\texttt{pfrac\_mul\_factorial}, \texttt{pfrac\_div\_factorial}, and
the sqrt-rational split round-trip across a sweep of arguments,
and verify Legendre's formula~\cref{eq:legendre} against a direct
factorial computation up to $n\le 100$.

\paragraph{Out-of-memory handling}
A dedicated harness \texttt{test\_oom.c} forces the $(N+1)$-th
allocation in each public symbol path to fail and verifies that the
library aborts cleanly with \texttt{SIGABRT}, i.e.\ that the
library reaches the diagnostic-and-abort path in \texttt{xalloc.c}
rather than dereferencing a null pointer. Each forced failure is run
in a forked child process so that the parent's state remains clean
across the sweep over $N$. This guards against silent corruption
under memory pressure.

\paragraph{Continuous integration}
A pair of CI workflows exercises the library on every push and
pull request: a GitHub Actions workflow covering the cells whose
hosted runners are unmetered for public repositories, and a
CircleCI workflow covering the native arm64 Linux cell, a
musl-libc x86-64 cell on Alpine, and a 32-bit i686 cell on Debian
that have no GitHub-hosted equivalent.  A source-list consistency
check guards against drift between the CMake build and the Python
\texttt{setup.py}.  The combined coverage spans the following
dimensions:
(i)~Linux/GCC (shared and static), Linux/Clang, macOS/Clang, and
Windows/MSVC (x86-64 and arm64), with the optional Fortran,
libquadmath, and MPFR back-ends enabled on Linux and a baseline
no-optional-features cell;
(ii)~a dedicated cell with the optional FLINT bigint back-end
enabled to exercise the \texttt{fmpz\_t} arithmetic path of
\cref{sec:limits};
(iii)~the source rebuilt on Linux/Clang under the address and
undefined-behaviour sanitizers
(\texttt{-fsanitize=\allowbreak address,\allowbreak undefined,\allowbreak -fno-sanitize-recover=all}),
so any out-of-bounds access or unsigned-overflow bug fails the
build;
(iv)~the Python extension built and tested under three CPython
versions (3.9, 3.11, 3.13) on Linux and Windows;
(v)~four override cells that smoke-test alternative dispatch arms
on x86-64: \texttt{BIGINT\_FORCE\_PORTABLE} (pure-C99 fallback),
\texttt{BIGINT\_NO\_DIVQ} (no inline-asm \texttt{divq}),
\texttt{-fno-lto}, and a no-TLS fallback.  Each cell confirms that
the selected dispatch arm produces bit-identical output to the
default build;
(vi)~an out-of-tree \texttt{find\_package(wignernj)} downstream
test covering the four shared/static $\times$ with/without-Fortran
configurations;
(vii)~a code-coverage cell that runs the full ctest plus pytest
suite under \texttt{lcov} instrumentation and uploads the merged
\texttt{coverage.info} to Codecov, annotating every push and
pull request with line- and branch-coverage diffs.

\section{Benchmarks}
\label{sec:benchmarks}

To put \texttt{libwignernj}'s performance in context, we
benchmark the originally deposited version~0.5.0 against two
reference libraries; the plotted numbers remain representative
of the currently deposited v0.8.0.
\texttt{WIGXJPF}~1.13 by \citet{Johansson2016_SJSC_376} is the
canonical prime-factorisation implementation; the algorithm
published in that work is also the basis of the present
clean-room implementation. The GNU Scientific Library
(GSL~2.8)~\cite{GSLmanual} provides the routines
\texttt{gsl\_sf\_coupling\_3j}, \texttt{gsl\_sf\_coupling\_6j},
and \texttt{gsl\_sf\_coupling\_9j}, which are widely used and
written in pure double-precision floating point. GSL appears to
evaluate the 3j and 6j symbols by the Racah single-sum formula
and the 9j symbol through Wigner's expansion as a sum of
products of three 6j symbols. We benchmark \texttt{libwignernj}
in two configurations: with the default in-tree multiword
bigint kernel, and with the optional FLINT bigint back-end.
\texttt{sympy.physics.wigner}~\cite{Meurer2017_PCS_103} is not
included in the timing comparison, as it is a pure-Python
symbolic implementation rather than a compiled-language
library, and the per-evaluation cost is dominated by the
interpreted symbolic-arithmetic stack rather than by the
underlying algorithm.

The benchmark targets four representative inputs that span the
full range of behaviour: the 3j symbols
$\bigl(\begin{smallmatrix} j & j & j \\ j & -j & 0 \end{smallmatrix}\bigr)$
and
$\bigl(\begin{smallmatrix} j & j & j \\ 0 & 0 & 0 \end{smallmatrix}\bigr)$,
the all-equal-$j$ 6j $\{j\,j\,j;\,j\,j\,j\}$, and the
all-equal-$j$ 9j $\{j\,j\,j;\,j\,j\,j;\,j\,j\,j\}$.  The
$(j\,j\,j;\,0\,0\,0)$ and the all-equal-$j$ 9j vanish at odd $j$
(by the $j_1+j_2+j_3$ parity rule and the 9j row/column-swap
symmetry, respectively) and are evaluated at even $j$ only.
All four cases are run for $j = 1\ldots 200$ (with the parity
restrictions noted).  For each $(j, \mathrm{symbol},
\mathrm{library})$ we record both the value and the per-call
wall-clock time.  The value is compared against a reference
generated by evaluating each symbol as an exact rational with
\texttt{sympy.\allowbreak physics.\allowbreak
wigner}~\cite{Meurer2017_PCS_103} and then rounding the result
to 40 decimal digits ($\sim 133$ bits, well above the 113-bit
mantissa of IEEE~754 binary128) with
mpmath~\cite{Johansson_mpmath}, so that the reference itself
carries no rounding error visible at any precision exposed by
\texttt{libwignernj} including its libquadmath back-end.  The
per-call time is taken as the minimum over an adaptive inner
loop of repeats whose count is chosen so the loop runs for at
least tens of milliseconds.  All three libraries were built
from source with the same compiler (GCC 16.0.1) and the same compile
flags, namely the default Fedora optimization flags (queryable on a
Fedora system through \texttt{\$(rpm -E \%optflags)}).  The CPU was
a single core of a 12th-generation Intel Core i5-1235U (Alder Lake,
$4.4\,\mathrm{GHz}$ maximum single-core boost) running on Fedora~44
with the CPU frequency governor pinned to \texttt{performance}.
The complete benchmark harness, reference generator, and plotting
scripts are included as supplementary material to this paper.

\begin{figure*}[t]
\centering
\includegraphics[width=0.95\textwidth]{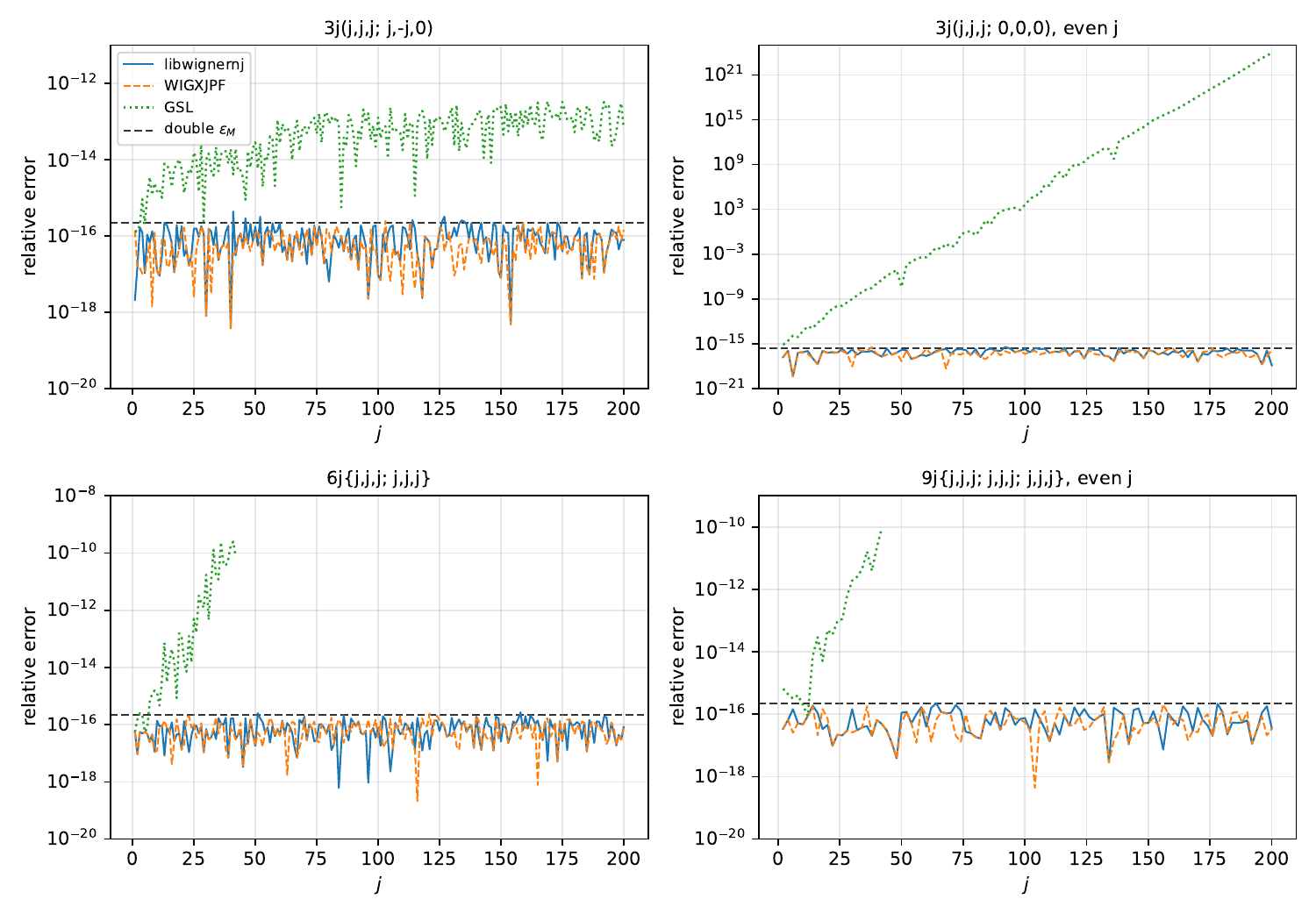}
\caption{Relative error of \texttt{libwignernj}, \texttt{WIGXJPF}~1.13,
  and GSL~2.8 against an mpmath quadruple-precision reference for
  the four benchmark inputs at $j = 1\ldots 200$.  The dashed
  black line marks the double-precision unit roundoff
  $\epsilon_M$.  GSL points are shown only for $j$ at which GSL
  returned a finite value without raising its internal error
  handler. Values where GSL trapped or returned a non-finite
  result are omitted from the plot.  \texttt{libwignernj} (default
  and FLINT back-ends produce bit-identical doubles, so a single
  curve) and \texttt{WIGXJPF} agree with the reference to the
  unit roundoff or below at every $j$, while GSL's
  gamma-function-based recursion silently loses many orders of
  magnitude beyond a relatively small $j$.}
\label{fig:accuracy}
\end{figure*}

\begin{figure*}[t]
\centering
\includegraphics[width=0.95\textwidth]{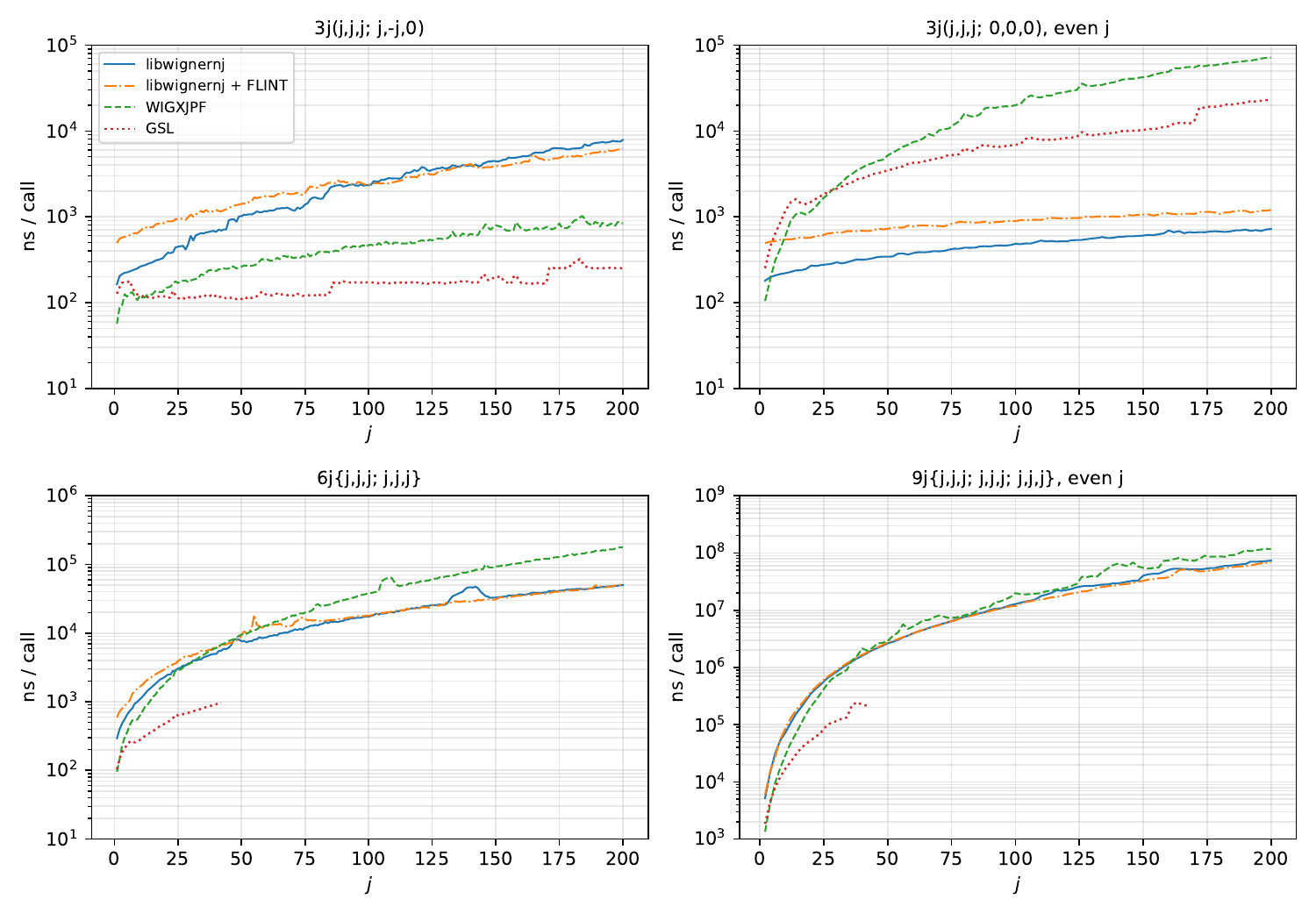}
\caption{Per-call wall time of \texttt{libwignernj} (in-tree multiword
  bigint kernel, solid. With the optional FLINT back-end,
  dash-dotted), \texttt{WIGXJPF}~1.13 (dashed), and GSL~2.8 (dotted)
  on the four benchmark inputs at $j = 1\ldots 200$.  The two 3j
  panels share the y-axis. The 6j and 9j panels are scaled
  independently.  GSL timing samples are dropped where GSL did not
  return a meaningful value (same mask as
  \cref{fig:accuracy}).}
\label{fig:timing}
\end{figure*}

\paragraph{Accuracy}
\Cref{fig:accuracy} shows that \texttt{libwignernj} and
\texttt{WIGXJPF} agree with the mpmath reference at the unit
roundoff or below across every $j$ in every panel---both
implementations sit on the dashed $\epsilon_M$ line or beneath it,
and at many $j$ the relative error is exactly zero (rendered at
the bottom of each panel by the floor used in the plot).  The
default and FLINT back-ends of \texttt{libwignernj} produce
bit-identical doubles at every $j$, verified independently in the
continuous-integration matrix, so a single \texttt{libwignernj}
trace appears in the accuracy plot.  GSL, by contrast, is correct to the unit roundoff only at small
$j$ and then degrades catastrophically as the Racah-sum
intermediates exceed the dynamic range of double-precision
floating point: the relative error climbs through many orders
of magnitude in every panel, reaching values far in excess of
unity. The $j$ values at which GSL trapped via its error
handler or returned a non-finite result are omitted from the
plot. This silent loss of accuracy without an error indication
is the principal motivation for the prime-factorisation scheme
that \texttt{libwignernj} adopts.

\paragraph{Timing}
\Cref{fig:timing} shows the per-call cost of each library on
the same four inputs, with a different picture in each panel.
GSL is the fastest where it produces correct output: a few
hundred nanoseconds for 3j and 6j at small $j$, with the curve
cut off past the $j$ at which the Racah-sum intermediates exceed
the dynamic range of double-precision floating point.  On the
$\bigl(\begin{smallmatrix}j&j&j\\0&0&0\end{smallmatrix}\bigr)$
panel \texttt{libwignernj} is the fastest above $j\approx 30$,
by the closed-form fast path of \cref{eq:3j-000-closed} that
skips the entire Racah-sum machinery and beats \texttt{WIGXJPF}
and GSL by roughly an order of magnitude at large $j$.  On the
$\bigl(\begin{smallmatrix}j&j&j\\j&-j&0\end{smallmatrix}\bigr)$
3j panel the situation is reversed: \texttt{WIGXJPF} pulls ahead
of \texttt{libwignernj} at large $j$ by roughly an order of
magnitude.  On the all-equal-$j$ 6j \texttt{libwignernj} is the
faster of the two, and the margin widens with $j$.  On the
all-equal-$j$ 9j the two libraries track each other closely
across the entire range.  FLINT~\cite{Hart2010__88} implements a full ladder of
sub-quadratic multiplication algorithms
(Karatsuba~\cite{Karatsuba1963_SPD_595},
Toom-Cook~\cite{Toom1963_SMD_714,Cook1966__}, and
Sch\"onhage--Strassen~\cite{Schoenhage1971_C_281}); even so, the
optional FLINT back-end tracks the in-tree back-end closely on
every panel across the $j$ range studied here. The default
in-tree back-end already includes a Karatsuba path with a
small-word-count schoolbook crossover, and the typical-$j$
workload sits below the regime where FLINT's heavier
algorithmic ladder amortises its overhead. We expose FLINT
primarily as a future escape valve for callers running well
past the default prime-table ceiling, where its asymptotic
complexity advantage will eventually dominate. A plausible
source of the remaining small-$j$ gap is the set of algorithmic
optimisations of the prime-power-exponent bookkeeping that
\texttt{WIGXJPF} implements and \texttt{libwignernj} does not,
although we have not studied the difference directly:
\texttt{libwignernj} is intentionally a clean-room
implementation derived from the published methods rather than
from \texttt{WIGXJPF}'s source.
The per-call \texttt{pfrac\_t} and LCM-exponent buffers are
themselves cached per-thread on first use, so they do not
contribute to the per-call cost (\cref{sec:software}).

The per-call timings exclude one-time memory-allocation costs
in both libraries; they reflect the steady-state cost of
evaluating one symbol with all per-thread buffers already warm
in the inner benchmark loop. \texttt{WIGXJPF} requires the
caller to allocate prime-decomposition tables and per-thread
scratch arrays through \texttt{wig\_\allowbreak
table\_\allowbreak init} and \texttt{wig\_\allowbreak
temp\_\allowbreak init} before any symbol is evaluated, with
allocation sizes growing with the maximum $2j$ to be passed.
\texttt{libwignernj} grows its per-thread scratch lazily on
first call and exposes a \texttt{wignernj\_warmup\_to} entry
point (\cref{sec:software}) that lets the caller pre-grow
both per-thread caches to fit the worst-case factorial argument
of an upcoming workload. The prime table itself is a
compile-time constant and incurs no run-time allocation in
either library. Excluding allocation from the timing sweep is
the standard methodology for libraries of this kind, since
allocation cost is amortised away in any realistic workload
that evaluates many symbols per allocated scratch.

\paragraph{Practical implications}
The performance picture is mixed: \texttt{libwignernj} is
faster than \texttt{WIGXJPF} on the
$(j\,j\,j;\,0\,0\,0)$ 3j and the all-equal-$j$ 6j, comparable
on the all-equal-$j$ 9j, and slower (by roughly an order of
magnitude at large $j$) on the
$\bigl(\begin{smallmatrix}j&j&j\\j&-j&0\end{smallmatrix}\bigr)$
3j.  GSL is faster than both where it produces correct output,
but its operating range is bounded by the dynamic range of
double-precision floating point in the Racah-sum
intermediates and so is unsuited for moderate-to-large $j$.
Two considerations make these trade-offs acceptable in
practice. First, the licence: GSL is distributed under the
GNU~GPL~v3 and \texttt{WIGXJPF} under the GNU~LGPL~v3; both are
copyleft and impose obligations on downstream users that the
permissive BSD~3-Clause licence of \texttt{libwignernj} does
not, making the library easy to embed in proprietary or
differently-licensed scientific software. Second, the
evaluation of Wigner symbols is rarely the bottleneck of an
end-to-end calculation: in atomic, molecular, and nuclear
applications the dominant cost is typically that of the radial
two-electron integrals, the matrix diagonalisation, or the
Monte-Carlo / molecular-dynamics outer loop, all of which take
orders of magnitude longer per step than the angular-momentum
algebra they consume. For applications that do require many
millions of coefficients in a hot inner loop, a complementary
use of \texttt{libwignernj} is to populate a lookup table once
at start-up: the table is correct to last bit by construction,
and a subsequent table look-up is essentially free. The compact
storage schemes of \citet{Rasch2004_SJSC_1416} or
\citet{Pinchon2007_IJQC_2186} can be used to keep the table
size tractable.

\section{Examples}
\label{sec:examples}

We illustrate the library with three short examples:
angular-momentum recoupling via the 6j symbol, a side-by-side
evaluation of the Gaunt coefficient over complex and real
spherical harmonics, and an extreme-angular-momentum case in
which the symbol's value lies below the smallest representable
double-precision number, so that the optional MPFR back-end is
needed.

\subsection{Recoupling of three angular momenta}

Given three angular momenta $j_1$, $j_2$, $j_3$ coupled to total $J$,
two equally valid coupling schemes are
\begin{align}
|((j_1 j_2) j_{12},\, j_3)\, JM\rangle  &\quad \text{and} \quad
|(j_1,\, (j_2 j_3) j_{23})\, JM\rangle.
\end{align}
The unitary transformation between them is given by the Wigner 6j symbol:
\begin{align}
&\langle ((j_1 j_2) j_{12},\, j_3)\, JM
   \,|\, (j_1, (j_2 j_3) j_{23}) JM \rangle \nonumber\\
&\quad= (-1)^{j_1+j_2+j_3+J}\,
       \sqrt{(2j_{12}+1)(2j_{23}+1)}
\begin{Bmatrix} j_1 & j_2 & j_{12} \\ j_3 & J & j_{23} \end{Bmatrix}.
\label{eq:recoupling}
\end{align}
The right-hand side is exactly the Racah $W$-coefficient up to a
square-root prefactor. The 30-line program below evaluates it for
a full set of $j_{12}$ and $j_{23}$ couplings, demonstrating that
$\sum_{j_{12}} \langle\cdots|\cdots\rangle^2 = 1$ (a unitary-frame
sum rule) is satisfied to last-bit accuracy:
\begin{lstlisting}[language=C]
#include <stdio.h>
#include <math.h>
#include "wignernj.h"

int main(void) {
    /* Couple three j=1/2 spins (tj = 1).
     * j12 ranges over 0, 1; j23 over 0, 1.
     * For fixed j23 = 1, sum_{j12} |<...|...>|^2 must equal 1. */
    int tj1 = 1, tj2 = 1, tj3 = 1, tJ = 1;
    int tj23 = 2;                       /* j23 = 1 */
    double sum = 0.0;
    for (int tj12 = 0; tj12 <= 2; tj12 += 2) {
        double w = racah_w(tj1, tj2, tJ, tj3, tj12, tj23);
        double pref = sqrt((tj12 + 1.0) * (tj23 + 1.0));
        double a = pref * w;
        printf("j12=%d  <(j1 j2)j12 j3 | j1 (j2 j3)j23> = %+.15f\n",
               tj12 / 2, a);
        sum += a * a;
    }
    printf("sum-of-squares = %.15f  (should be 1)\n", sum);
    return 0;
}
\end{lstlisting}
The program produces, on every supported platform,
\texttt{sum-of-squares = 1.000000000000000} to last-bit accuracy,
confirming the unitarity of the recoupling transformation.

\subsection{Complex- versus real-spherical-harmonic Gaunt coefficient}

For a generic set of indices both
$\mathcal{G}^C \equiv \int Y_{\ell_1}^{m_1} Y_{\ell_2}^{m_2}
Y_{\ell_3}^{m_3}\,\mathrm{d}\Omega$ and
$\mathcal{G}^R \equiv \int S_{\ell_1 m_1} S_{\ell_2 m_2}
S_{\ell_3 m_3}\,\mathrm{d}\Omega$ are non-zero, although their values
in general differ. The two are exposed by the library through
\texttt{gaunt} and \texttt{gaunt\_real} respectively. Taking
$(\ell_1,m_1,\ell_2,m_2,\ell_3,m_3)=(1,-1,1,-1,2,2)$, which satisfies
the selection rules of both routines, the evaluations
\begin{lstlisting}[language=C]
double gC = gaunt     (/*tl1=*/2, /*tm1=*/-2,
                       /*tl2=*/2, /*tm2=*/-2,
                       /*tl3=*/4, /*tm3=*/ 4);
double gR = gaunt_real(/*tl1=*/2, /*tm1=*/-2,
                       /*tl2=*/2, /*tm2=*/-2,
                       /*tl3=*/4, /*tm3=*/ 4);
\end{lstlisting}
return $\mathcal{G}^C = +0.30901936161855165$ and
$\mathcal{G}^R = -0.21850968611841581$, both correctly rounded to
the last bit. The same indices at long-double precision are
obtained by appending the suffix \texttt{\_l} to either function
name, and at user-chosen precision through \texttt{gaunt\_mpfr} or
\texttt{gaunt\_real\_mpfr}.

\subsection{An evaluation below the IEEE 754 underflow threshold}

For very large angular momenta, individual coupling coefficients
can have magnitudes far below the smallest representable
double-precision floating-point number ($\sim 5\times 10^{-324}$
for subnormals, $\sim 2.2\times 10^{-308}$ for normals).  A direct
floating-point evaluation would silently produce zero;
\texttt{libwignernj}'s prime-factorisation pipeline avoids that
fate because all intermediate arithmetic is exact and underflow
can occur only at the single final cast.  The 80-bit
\texttt{long double} on x86-64, the libquadmath binary128
back-end, and the MPFR back-end each push the underflow boundary
out by hundreds or thousands of orders of magnitude.

Consider the 3j symbol
$\bigl(\begin{smallmatrix}4000 & 4000 & 4000 \\ 3995 & -3995 & 0\end{smallmatrix}\bigr)$.
Its true magnitude is $\sim 8\times 10^{-443}$, well below the
double-precision floor. A short program that evaluates it across
the available precisions is
\begin{lstlisting}[language=C]
#include <stdio.h>
#include <quadmath.h>
#include <mpfr.h>
#include "wignernj.h"
#include "wignernj_quadmath.h"
#include "wignernj_mpfr.h"

int main(void) {
    int tj = 2 * 4000;            /* 2*j */
    int tm = 2 * 3995;            /* 2*m */
    double      d = wigner3j  (tj, tj, tj, tm, -tm, 0);
    long double l = wigner3j_l(tj, tj, tj, tm, -tm, 0);
    __float128  q = wigner3j_q(tj, tj, tj, tm, -tm, 0);
    char buf[64]; quadmath_snprintf(buf, sizeof buf, "%.30Qe", q);
    mpfr_t v;
    mpfr_init2(v, 256);           /* 256-bit precision */
    wigner3j_mpfr(v, tj, tj, tj, tm, -tm, 0, MPFR_RNDN);
    printf("double      : %.16e\n",  d);
    printf("long double : %.20Le\n", l);
    printf("__float128  : %s\n",     buf);
    printf("MPFR (256b) : ");
    mpfr_printf("%.30Re\n", v);
    mpfr_clear(v);
    return 0;
}
\end{lstlisting}
which prints
\begin{verbatim}
double      : 0.0000000000000000e+00
long double : 8.42709741643091937108e-443
__float128  : 8.427097416430919371371619698913e-443
MPFR (256b) : 8.427097416430919371371619698913e-443
\end{verbatim}
The double-precision result has underflowed to zero. The
long-double result is the correctly-rounded $\sim 80$-bit
approximation. The binary128 result agrees with MPFR through its
113-bit mantissa.  The MPFR result provides as many additional
digits as the caller-chosen precision permits.

\section{Availability and conclusion}
\label{sec:conclusion}

\texttt{libwignernj} is freely available under the BSD~3-Clause
licence from \url{https://github.com/susilehtola/libwignernj},
including the C, C++, Fortran 90 and Python interfaces, the
optional libquadmath and MPFR back-ends, the optional FLINT bigint
back-end, the test suite, and the supporting tooling (the
prime-table generator and the \texttt{sympy}-based reference
generator). Pre-built Python wheels for Linux, macOS, and Windows
are published on PyPI and installable via \texttt{pip install
wignernj}. The library is self-contained C99, has no mandatory
external dependencies, and provides exact, last-bit-correct
evaluation of Wigner 3j, 6j and 9j symbols, Clebsch--Gordan
coefficients, Racah $W$ coefficients, Fano $X$-coefficients, and
Gaunt coefficients over both complex and real spherical harmonics,
across the full range of angular momenta of practical interest.

The source code combines the prime-factorisation technique of
\citet{Dodds1972_CPC_268} for the integer arithmetic with the
multiword-integer Racah sum of \citet{Johansson2016_SJSC_376}. The
new contributions of the present work are (i)~the assembly of the
entire algorithm into a small, portable C99 library with no
caller-side initialisation and clean C, C++, Fortran~90, and
Python bindings; (ii)~the absorption of Clebsch--Gordan, Racah
$W$, Fano $X$, and both complex and real-spherical-harmonic Gaunt
coefficients into the same exact pipeline as the 3j/6j/9j;
(iii)~the optional libquadmath and MPFR back-ends that re-use the
same exact tuple to deliver IEEE~754 binary128 and
arbitrary-precision results at the user's chosen precision;
(iv)~the systematic verification, sanitiser, and out-of-memory
testing of \cref{sec:verification}, exercised by a
multi-platform continuous-integration pipeline that builds and
runs the entire test suite on every push and pull request; and
(v)~a build system that ships an installable CMake package and
a \texttt{pkg-config} file, so that \texttt{libwignernj} can be
dropped into a larger CMake stack either as an installed
dependency or as a git submodule, exposing the same target
namespace in both cases---a pattern that, to the author's
knowledge, none of the earlier public implementations supports
out of the box.

\paragraph{Position relative to the open-source ecosystem}
\texttt{libwignernj} shares the bit-exact prime-factorisation
pipeline of the Johansson--Forss\'en
family (\texttt{WIGXJPF}~\cite{Johansson2016_SJSC_376},
\texttt{WignerSymbols.jl}~\cite{Haegeman_WignerSymbolsjl},
\texttt{wigners}~\cite{Fraux_wigners}) but distinguishes itself on
three axes simultaneously: (i)~it is self-contained C99 with no
caller-side initialisation, embeddable from C, C\texttt{++},
Fortran 90, and Python through a common public application binary
interface (ABI, and from any other language with a C
foreign-function interface);
(ii)~it has optional libquadmath and GNU~MPFR back-ends sharing
the same exact intermediate tuple, so the caller can request
additional precision without changing algorithm.  Finally (iii) it
extends the same exact pipeline to Clebsch--Gordan, Racah $W$,
Fano $X$, and both complex and real-spherical-harmonic Gaunt
coefficients within a single library.  None of the libraries in
\cref{tab:oss} currently combines all three properties, so
they are best viewed as a complementary set.

The permissive BSD~3-Clause licence of \texttt{libwignernj} was
chosen to attract users whose own projects cannot accept the
copyleft
obligations of the GNU Scientific Library or \texttt{WIGXJPF} and
have so far had to reimplement the angular-momentum machinery
in-house. We hope that this reusable implementation will be
widely adopted by the community.

\paragraph{Future outlook}
The deliberately small, focused, reusable design means that any
future algorithmic refinement to the shared C core is inherited
automatically by every language binding. The
continuous-integration pipeline of \cref{sec:verification}
provides a high-confidence harness against which contributors
can validate such refinements before they land upstream. Two
natural directions stand out. First, an extension beyond the 9j
symbol: \texttt{libwignernj} stops at the 9j, in line with
every implementation in \cref{tab:prior-work,tab:oss} that
goes beyond 6j, since 9j is the largest 3$n$-$j$ that is
uniquely defined (at $n=4$ two distinct topologies appear, and
the count of inequivalent 3$n$-$j$ topologies grows further at
higher $n$). Nothing in the prime-factorisation pipeline is
specific to the 9j recursion depth, however: the same exact
tuple (\cref{eq:exact-tuple}) and the same multiword Racah-sum
machinery extend immediately to either 12j by accumulating four
or three exact-tuple products per term in the outer summation,
just as the 9j does with three. The barriers are pragmatic
rather than algorithmic---demand is small relative to the 9j,
and there is no widely-available reference implementation to
validate against---so a 12j extension is left to a future
release on demand.

Second, graphics processing units (GPUs) have become an
essential consideration in modern scientific computing, so it
is worth asking whether \texttt{libwignernj} can productively
exploit them. The hot path is multiword integer arithmetic on
\texttt{bigint\_t} operands of data-dependent length, the
access pattern that single-instruction multiple-thread (SIMT)
architectures penalise most heavily through divergent control
flow and irregular carry propagation. A bit-exact GPU port
would therefore deliver essentially serial per-thread
performance, while a floating-point port would re-implement the
recursion-based algorithms already in the libraries of
\cref{tab:oss} and lose the property that distinguishes the
present library. Callers who need batch throughput rather than
bit-exact individual evaluations are better served by CPU
thread-level parallelism over the per-thread caches of
\cref{sec:software}, or by one of the recursion-based
libraries. The symbols can also be precomputed on the CPU and
uploaded to a lookup buffer on the GPU, letting the GPU kernel
consume the exact values without performing the symbol
evaluation itself.

\section*{CRediT authorship contribution statement}
\textbf{Susi Lehtola}: Conceptualization, Methodology, Software,
Validation, Formal analysis, Investigation, Writing---original
draft, Writing---review and editing, Visualization, Funding
acquisition.

\section*{Declaration of competing interest}
The author declares that he has no known competing financial
interests or personal relationships that could have appeared to
influence the work reported in this paper.

\section*{Data availability}
No new data were generated or analysed in the course of this work.
The source code, build scripts, and test suite described above are
openly available under the BSD 3-Clause licence at
\url{https://github.com/susilehtola/libwignernj}. The verification
reference tables (\cref{sec:verification}) are regenerated
deterministically by the
\texttt{tests/\allowbreak gen\_\allowbreak refs.py} script that
ships with the source repository.

\section*{Supporting Information}
The supporting information accompanying this paper contains the
material needed to reproduce the benchmark results of
\cref{sec:benchmarks}, none of which is part of the main
\texttt{libwignernj} repository because it would otherwise pull
in build-time dependencies (\texttt{WIGXJPF} and GSL) that the
library itself does not require.  Specifically, it provides:
(i)~the comparative benchmark harness that links against
\texttt{libwignernj}, \texttt{WIGXJPF}~1.13, and GSL~2.8 and
produces the per-call timings of \cref{fig:timing};
(ii)~the accuracy-comparison driver and \texttt{mpmath} reference
generator that produce the relative-error data of
\cref{fig:accuracy}; and (iii)~the plotting scripts that
generate both figures.  All scripts are released under the same
BSD 3-Clause licence as the library and have been tested under
the build configuration documented in \cref{sec:benchmarks}.

\section*{Declaration of generative AI and AI-assisted technologies in the writing process}
During the preparation of this work the author used Anthropic's
Claude (model: Opus~4.7, accessed via the Claude Code
command-line interface) to
assist with drafting and revising sections of the manuscript, with
surveying the implementation literature against Crossref records,
with prototyping the build-system configuration and the
continuous-integration matrix, with generating the TikZ source of
the graphical abstract, and with code refactoring---notably the
prime-table-iteration optimisation of \cref{sec:limits}.
After using this tool, the author reviewed and edited the content
as needed and takes full responsibility for the contents of the
publication.

\section*{Acknowledgments}
The author thanks the Academy of Finland for financial support under
project no.~350282 and 353749.

%--------------------------------------------------------------------
\bibliographystyle{elsarticle-num-names}
\bibliography{citations,software}

\end{document}